\RequirePackage{ifpdf}
\documentclass[hyper,letterpaper]{JHEP3}
\usepackage{amsmath,amssymb,amsfonts}
\usepackage{fancybox}
\usepackage{cite}
\usepackage{graphicx, wrapfig}
\usepackage{verbatim}
\usepackage{appendix}
\DeclareRobustCommand{\quantumbinomial}{\genfrac{[}{]}{0pt}{}}

\title{Junctions of refined Wilson lines and one-parameter deformation of quantum groups.}

\author{Sungbong Chun \\
Walter Burke Institute for Theoretical Physics, California Institute of Technology, Pasadena, CA 91125 USA}

\abstract{We study junctions of Wilson lines in refined SU(N) Chern-Simons theory and their local relations. We focus on junctions of Wilson lines in antisymmetric and symmetric powers of the fundamental representation and propose a set of local relations which realize one-parameter deformations of quantum groups $\dot{U}_{q}(\mathfrak{sl}_{m})$ and $\dot{U}_{q}(\mathfrak{sl}_{n|m})$.
\\
\\
\\
\\
\\
\\
\\
{\tt CALT-TH-2017-002}}

\begin{document}
\cornersize{1}

\section{Introduction}
\label{sec:intro}
Chern-Simons gauge theory is a 3d TQFT which has long served as a bridge between quantum physics and knot theory. Its gauge invariant observables, the Wilson loop operators, are supported on knots/links $K \subset S^{3}$. When the gauge group is $SU(N)$, their expectation values are equal to $\mathfrak{sl}_{N}$ knot polynomials of $K$ \cite{Witten89}. The correspondence extends to the spectrum of BPS states and the homological invariants of knots. $\mathfrak{sl}_{N}$ knot polynomials can be ``categorified'' to $\mathfrak{sl}_{N}$ knot homologies in a sense that the latter has the former as its graded dimension \cite{Kh, KhR1, KhR2, Wu}. Physically, the expectation value of a Wilson loop operator can be categorified to the spectrum of BPS states in a configuration of intersecting M5 branes. The former is equal to the graded dimension of the latter \cite{GV1, GV2, GV3, OV, LMV}, and the spectrum of BPS states themselves realize the $\mathfrak{sl}_{N}$ knot homologies \cite{GSV,Gukov2007,Witten2011}.

\begin{figure} [htb]
\centering
\includegraphics{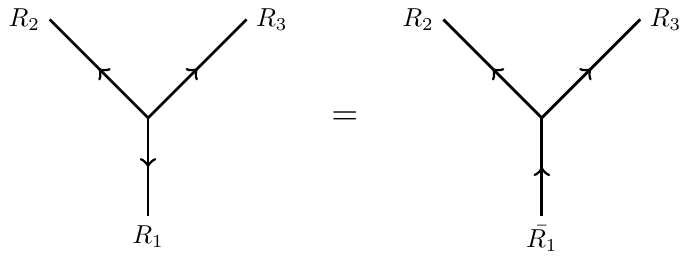} 
\caption{LHS: a junctions of three Wilson lines colored in $R_{1}, R_{2}, R_{3}$ such that $0 \in R_{1} \otimes R_{2} \otimes R_{3}$. At the junction, we place a gauge invariant tensor in $Hom_{G}(R_{1} \otimes R_{2} \otimes R_{3}, \mathbb{C})$. RHS: an equivalent junction, with $R_{1}$-strand reversed and replaced by its complex dual.}
\label{fig:MOYjunction}
\end{figure}

\begin{figure} [htb]
\centering
\includegraphics{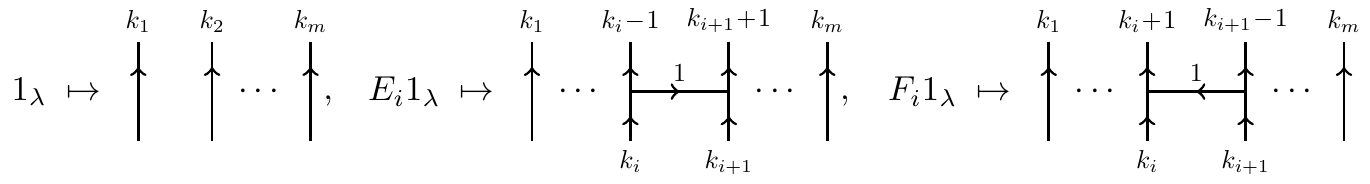}
\caption{Skew-Howe duality functor of \cite{CKM}: a correspondence between the generators of (idempotented) quantum group $\dot{\mathbf{U}}_{q}(\mathfrak{sl}_{m})$ and network of Wilson lines. Here, $\lambda = (k_{1}, \cdots, k_{m})$ is a $\mathfrak{sl}_{m}$ weight, and $k_{i}$ stands for $\wedge^{k_{i}}\square$, the $k_{i}$-th antisymmetric representation of SU(N).}
\label{fig:skewHowe}
\end{figure}

We can also introduce junctions of Wilson lines as in Figure \ref{fig:MOYjunction}. By placing a gauge invariant tensor at each junction, we can define a gauge invariant observable for every trivalent graph in $S^{3}$ \cite{Witten89wf}. Upon path integral, a network of Wilson lines in a 3-manifold $M_{3}$ with punctured boundaries fix a vector in the associated Hilbert space $H_{\{ \partial M_{3}; R_{1},\cdots,R_{n} \}}$, where $\{R_{i} \}$ are the representation of Wilson lines crossing the boundary. When the Hilbert space $H_{\{ \partial M_{3}; R_{1},\cdots ,R_{n} \}}$ is finite-dimensional, sufficiently many Wilson lines sharing the same boundary condition $\{ \partial M_{3}; R_{1},\cdots ,R_{n} \}$ would satisfy a linear relation. In \cite{Witten89rw}, networks of Wilson lines in spin representations (of $SU(N)$) were shown to satify a set of linear relations which can be identified with the generating relations of a quantum group. One can repeat the same procedure for networks of Wilson lines in antisymmetric representations, and a set of linear relations among them realize the generating relations of (idempotented) quantum groups $\dot{\mathbf{U}}_{q}(\mathfrak{sl}_{m})$ \cite{CGR}. Below, we provide networks of Wilson lines which correspond to the generators of $\dot{\mathbf{U}}_{q}(\mathfrak{sl}_{m})$ (Figure \ref{fig:skewHowe}.) They are ``generators'' in a sense that their multiplication (vertical stacking) and addition (formal sum) generate the entire $\dot{\mathbf{U}}_{q}(\mathfrak{sl}_{m})$ as an algebra. As generators of the quantum group, they also satisfy the generating relations of $\dot{\mathbf{U}}_{q}(\mathfrak{sl}_{m})$, which can be found in Equation \ref{eqn:uqslm}.
\begin{equation}
\begin{split}
1_{\lambda}1_{\lambda '} = \delta_{\lambda, \lambda ' }1_{\lambda}, \quad  E_{i}1_{\lambda} =  1_{\lambda+l_i} E_{i}, \quad F_{i}1_{\lambda}=1_{\lambda-l_i}F_{i}, \\[1.5ex]
[E_{i},F_{j}]1_{\lambda} = \delta_{i,j}[\lambda_{i}]1_{\lambda}, \quad [E_{i},E_{j}]1_{\lambda} = 0 \quad \text{for} \quad |i-j|>1, \\[1.5ex]
\text{and} \quad E_{i}E_{j}E_{i}1_{\lambda} = E_{i}^{(2)}E_{j}1_{\lambda}+E_{j}E_{i}^{(2)}1_{\lambda} \quad \text{for} \quad |i-j|=1,
\end{split}
\label{eqn:uqslm}
\end{equation}
where $l_{i} = (0, \cdots, 0, -1,1,0, \cdots, 0)$ in the same basis as $\lambda = (k_{1}, \cdots, k_{m})$, and $-1$ appears in the $i$-th position. 

In this paper, we generalize the above correspondence betwen networks of Wilson lines and quantum groups to the junctions in refined Chern-Simons theory. When Chern-Simons theory is defined on a Seifert manifold, its embedding into a five-brane configuration has an extra $U(1)_{\beta}$ flavor symmetry, and the expectation values of Wilson loops are ``refined'' by a one-parameter deformation \cite{AS}. Networks of Wilson lines can also be embedded in five-brane configurations \cite{CGR}, which also admit an extra flavor symmetry. As a result, local relations of Wilson lines are refined to the $\beta$-deformed quantum group $\dot{\mathbf{U}}_{q_{1},q_{2}}(\mathfrak{sl}_{m})$ relations, which are presented in two forms, one with twisted commutators and the other with ordinary commutators. For the networks of refined Wilson lines, however, the normalization ambiguities of the junctions \cite{Witten89wf} are not yet fully understood, and we can only suppose a particular choice of the normalization.

\section{Review: junctions in ordinary Chern-Simons theory}
\label{sec:review}
Let us briefly review here the junctions in ordinary Chern-Simons theory \cite{Witten89wf} and their embeddings in a five-brane configuration \cite{CGR}. 

\subsection{Expectation values and local relations of Wilson lines}
\label{subsec:expvalue}
\begin{figure} [htb]
\centering
\includegraphics{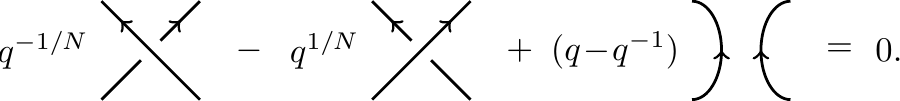}
\caption{Skein relation in ordinary $SU(N)$ Chern-Simons theory. The Wilson lines are lying in a closed 3-ball, vertically framed, and colored in $\square$, the fundamental representation. Here, $q = e^{\pi i / (N+k)}$.}
\label{fig:skein}
\end{figure}

Consider Chern-Simons theory on a 3-manifold $M_{3}$ with a gauge group $G$ and level $k$. A Wilson loop operator is a gauge invariant observable, defined by taking a holonomy of the gauge field $A$ along a prescribed loop $C$ and then taking a trace in a $SU(N)$ representation $R$. It is necessary to specify the ``framing'' of the Wilson loop as well, and we choose a vertical framing throughout this paper.
\begin{equation}
W_{R}(C) = \mathrm{Tr}_{R} \int_{C} A.
\label{eqn:Wilsonloop}
\end{equation}
When $G=SU(N)$, $M_{3}=S^{3}$, and $R=\square$ (the fundamental representation), we can easily compute the expectation values of Wilson loop operators by repeatedly using the skein relation (Figure \ref{fig:skein}) and the expectation value of an unknot \cite{Witten89}. 

Each term of Figure \ref{fig:skein} lies in a closed 3-ball, and the ends of Wilson lines meet the boundary $S^{2}$ (so there are four punctures on $S^{2}$, colored by $\square,\square,\bar{\square},\bar{\square}$.) A path integral on the closed 3-ball fixes a vector in the associated Hilbert space $H_{\{ S^{2};\square,\square,\bar{\square},\bar{\square} \}}$. The Hilbert space is isomorphic to the space of conformal blocks of $\widehat{\mathfrak{su}(N)}_{k}$ WZW model on $\{ S^{2};\square,\square,\bar{\square},\bar{\square}\}$, which is two-dimensional. Therefore, the skein relation represents a linear relation of three vectors in the two-dimensional vector space $H_{\{ S^{2};\square,\square,\bar{\square},\bar{\square} \} }$,
\begin{equation}
q^{-1/N} |\phi \rangle - q^{1/N} |\phi' \rangle + (q-q^{-1}) |\phi'' \rangle = 0,
\label{eqn:skein}
\end{equation}
where $|\phi \rangle, |\phi' \rangle, |\phi'' \rangle \in H_{\{ S^{2}; \square,\square,\bar{\square},\bar{\square} \} }$ are the vectors fixed by performing a path integral on Wilson line configurations of Figure \ref{fig:skein}. The coefficients are determined from the eigenvalues of the ``half-monodromy'' action on the Hilbert space.

\begin{figure} [htb]
\centering
\includegraphics{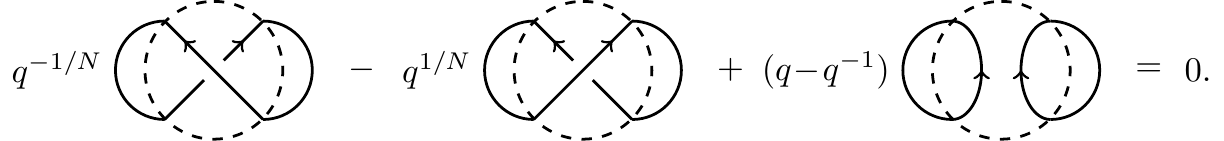}
\caption{A ``braid closure'' of Wilson lines involved in a skein relation. Again, the Wilson lines are colored in the fundamental representation and vertically framed.}
\label{fig:skein_closure}
\end{figure}

The skein relation is ``local'', in a sense that it holds upon cutting and gluing. For instance, let us glue the Wilson lines in Figure \ref{fig:skein} (inside the dashed circle of Figure \ref{fig:skein_closure}) with another Wilson line configuration in $S^{3}\setminus B^{3}$ (outside the dashed circle), along the boundary $S^{2}$ (dashed circle.)
From a vantage point of 3d TQFT, such a ``gluing'' action is equivalent to taking inner products of vectors. Let $\langle \psi |$ represent a vector fixed by a path integral outside the dashed circle of Figure \ref{fig:skein_closure}. Since the gluing in Figure \ref{fig:skein_closure} is simply an orientation-reversed identity morphism along the boundary $S^{2}$, $\langle \psi |$ lives in the dual space of $H_{\{ S^{2};\square,\square,\bar{\square},\bar{\square} \}}$. Take inner products of $\langle \psi |$ with $| \phi \rangle, | \phi' \rangle, | \phi'' \rangle$, and each inner product corresponds to a partition function $Z(S^{3},K)$ where $K$ is one of the Wilson loops in Figure \ref{fig:skein_closure}. Then, Figure \ref{fig:skein_closure} corresponds to an equation:
\begin{equation}
q^{-1/N} \langle \psi |\phi \rangle - q^{1/N} \langle \psi |\phi' \rangle + (q-q^{-1}) \langle \psi |\phi'' \rangle = 0.
\label{eqn:skein_closure}
\end{equation}
It is by this ``locality'' of skein relations that we can combinatorially compute the expectation value  $\langle W_{\square}(C) \rangle $ for a given knot $C \subset S^{3}$. Fix a projection of $C$, and repeatedly apply skein relations until all the crossings are resolved. Then, $C$ is written as a linear sum of disjoint unions of unknots in $S^{3}$, whose expectation values are already known.

\subsection{Wilson lines with junctions and their relations}
\label{subsec:junction}
As was briefly explained in the beginning, we can introduce junctions of Wilson lines and define gauge invariant observables supported on them. Consider a junction of $n$ Wilson lines colored in $R_{1}, \cdots, R_{n}$. Place a gauge invariant tensor $\epsilon \in Hom_{G}(R_{1} \otimes  \cdots \otimes R_{n}, \mathbb{C})$ at the junction. Given a closed graph of Wilson lines, we can contract the representation indices of Wilson lines with the gauge invariant tensors placed at each junctions and define a gauge invariant observable supported on the closed graph \cite{Witten89wf}. 

\begin{figure} [htb]
\centering
(circle removal) \quad \raisebox{-0.5\height}{\includegraphics{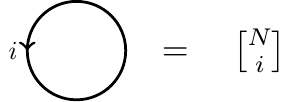}} \\[1.5ex]
(associativity) \quad \raisebox{-0.5\height}{\includegraphics{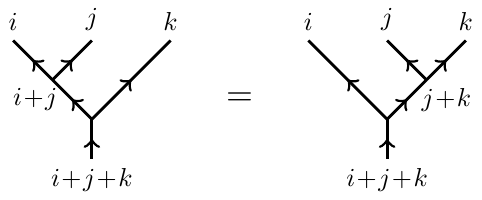}} \\[1.5ex]
(digon removal) \quad  \raisebox{-0.5\height}{\includegraphics{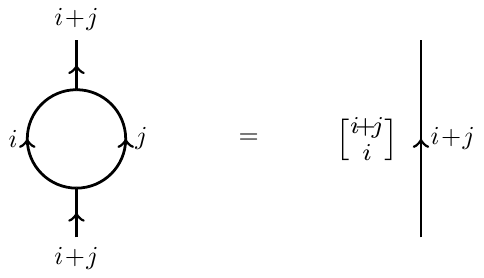}} \\[1.5ex]
($[E,F]$ relation) \quad \raisebox{-0.5\height}{\includegraphics{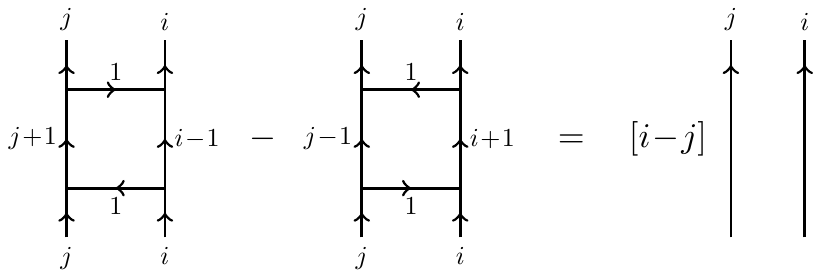}}
\caption{Local relations for the networks of Wilson lines in antisymmetric representations. $i,j,k$ are integers between $0$ and $N$, and our convention is that the integer labels $i,j,k$ represent the antisymmetric powers of the fundamental representation, $\wedge^{i}\square, \wedge^{j}\square, \wedge^{k}\square$. $h_{R}$ represents the conformal weight of the corresponding primary field in $\widehat{su(N)}_{k}$ WZW model}
\label{fig:relsPlanar}
\end{figure}

Now, consider a closed 3-ball around a network of Wilson lines, which meets the boundary $S^{2}$ at punctures $R_{1},\cdots,R_{n}$. Performing a path integral, we can fix a vector in the associated Hilbert space $H_{\{S^{2};R_{1},\cdots,R_{n}\}}$. The dimension of $H_{\{S^{2};R_{1},\cdots,R_{n}\}}$ is equal to the dimension of the $G$-invariant subspace of $R_{1} \otimes \cdots \otimes R_{n}$, which is finite-dimensional. When multiple networks of Wilson lines agree on the boundary $\{S^{2};R_{1},\cdots,R_{n}\}$, they define vectors in the same Hilbert space, and sufficiently many of them would satisfy a linear relation. Just like the skein relation, it is ``local'' and holds upon cutting and gluing. In general, its coefficients depend on normalizations of the gauge invariant tensors. In \cite{CGR}, a particular normalization was chosen for the trivalent junctions of Wilson lines in antisymmetric representations, so that they satisfy the relations in Figure \ref{fig:relsPlanar}. Here, we adopt a notation for the quantum binomials of $q$ for convenience: 
\begin{equation}
[i] = \dfrac{q^{i}-q^{-i}}{q-q^{-1}}, \quad [i]! = [i][i-1]\cdots [1], \quad \quantumbinomial{i+j}{i} = \dfrac{[i+j]!}{[i]![j]!}.
\label{eqn:q-binom}
\end{equation}

\subsection{Why junctions? Computability, MOY graph polynomials, and quantum groups}
\label{subsec:significance}
In fact, the normalization of junctions in Figure \ref{fig:relsPlanar} are chosen such that they realize the ``MOY graph relations'' and ``$N$Web category'', which appeared earlier in the knot theory literatures, \cite{MOY} and \cite{CKM} respecitvely. Wilson lines in Figure \ref{fig:relsPlanar} can be glued into closed planar trivalent graphs, whose edges are colored by integers between $0$ and $N$. These are exactly the ``MOY graphs'' appeared in \cite{MOY}, and the relations in Figure \ref{fig:relsPlanar} uniquely determine their ``MOY graph polynomials.'' Moreover, as long as the coloring of end points agree, a crossing can be written as a linear sum of MOY graphs, and we can write each colored $\mathfrak{sl}_{N}$ knot polynomial as a linear sum of MOY graph polynomials. 

In Chern-Simons theory, the resolution of crossings via MOY graphs corresponds to writing a Wilson loop in an antisymmetric representation as a linear sum of Wilson line networks in antisymmetric representations. Such technique can be particularly useful for computing the expectation values of Wilson loops colored in antisymmetric representations $\wedge^{i} \square$ for $i > 1$. The generalized skein relation for braided Wilson lines in $\wedge^{i}\square$ will involve higher twists in general, so it will be difficult to reduce the number of crossings by applying the generalized skein relations. On the contrary, we can always write the Wilson loop operator as a linear sum of Wilson line networks, and then we can apply local relations in Figure \ref{fig:relsPlanar} in a way that the number of junctions decreases.

Besides the advantages in computability, relations in Figure \ref{fig:relsPlanar} generate the defining relations of $\dot{\mathbf{U}}_{q}(\mathfrak{sl}_{m})$ (Equation \ref{eqn:uqslm}) under the skew-Howe duality functor (Figure \ref{fig:skewHowe}) \cite{CKM}. The quantum groups of interest are one-parameter deformations of the universal enveloping algebra of a classical Lie group \cite{BLM, Lus}, and they can be explicitly written as in Equation \ref{eqn:uqslm}. We can view the skew-Howe duality functor in Figure \ref{fig:skewHowe} as the networks of Wilson lines in antisymmetric representations comprising a representation of quantum groups $\dot{\mathbf{U}}_{q}(\mathfrak{sl}_{m})$. This is how we connect networks of Wilson lines, knot polynomials, and the representation theory of quantum groups. 

\subsection{Junctions in a five-brane configuration}
\label{subsec:fivebrane}
The correspondence among networks of Wilson lines, knot polynomials, and the representation theory of quantum groups extends to their categorifications \cite{CGR}. $\mathfrak{sl}_{N}$ knot homologies are defined from a homotopy category of matrix factorizations, and their constructions became fully combinatorial in the context of the representation theory of categorified quantum groups \cite{QR,BN}. Physical realization of $\mathfrak{sl}_{N}$ knot homologies were proposed in \cite{GW,CM,GS, CGR}, connecting the matrix factorizations of $\mathfrak{sl}_{N}$ homologies with those from the topological defects in Landau-Ginzburg models \cite{BHLS, KL, BR, HW}. In particular, the combinatorial construction of \cite{QR} is based on seamed surfaces and the associated matrix factorizations, which can be reproduced from a five-brane configuration of intersecting branes \cite{CGR}:
\begin{align}
\text{Space-time} &: \quad \mathbb{R}_{t} \times T^{*}M_{3} \times M_{4} \nonumber \\
N \, \text{M5-branes} &: \quad \mathbb{R}_{t} \times M_{3} \times D \label{eqn:gamma_fivebrane} \\
|R| \, \text{M5'-branes} &: \quad \mathbb{R}_{t} \times L_{\Gamma} \times D \nonumber
\end{align}
Above, $M_{3}$ is a 3-manifold, which will be assumed to be $S^{3}$ throughout this paper. $D \cong \mathbb{R}^{2} \subset M_{4} = TN$ is the ``cigar'' in the Taub-Nut space, and $\Gamma$ represents a trivalent junction of Wilson lines, colored in $SU(N)$ representations $R, R', R''$ such that $R = R' \otimes R''$. Lastly, $L_{\Gamma}$ is a Lagrangian submanifold of $T^{*}M_{3}$ intersecting $M_{3}$ along $\Gamma$. The above brane construction has $U(1)_{P} \times U(1)_{F}$ symmetry, where each factors are the rotation symmetries of $D$ and its normal bundle $D \subset M_{4}$, respectively.

The seamed surface $\mathbb{R}_{t} \times \Gamma$ is precisely the kind of seamed surfaces that appear in \cite{QR}. Although the given five-brane configuration is ``static'' in the ``time'' direction $\mathbb{R}_{t}$, we need to allow the seamed surfaces to fuse/split along the time direction as well, because the cobordisms in time direction correspond to the differentials of $\mathfrak{sl}_{N}$ knot homology in \cite{QR}. The whole five-brane setup preserves (at least) two real supercharges, and we can use them to obtain a 2d Landau-Ginzburg model on each facet of the seamed surface. B-type defects live on the junctions of surfaces, and their matrix factorization descriptions precisely coincide with those of \cite{BR, QR}. 

\section{Compatibility of junctions and refinement}
\label{sec:compatible}
Now we are ready to refine the junctions. Recall that each junction of Wilson lines comes with a gauge invariant tensor. We cannot immediately refine the gauge invariant tensor itself, because the gauge theory description of refined Chern-Simons theory is yet unknown. Thus, it is necessary to study the refinement of junctions from their embeddings in five-brane configurations (Equation \ref{eqn:gamma_fivebrane}.) Let us reproduce a five-brane setup of \cite{AS} here:

\begin{align}
\text{Space-time} &: \quad \mathbb{R}_{t} \times T^{*}M_{3} \times M_{4} \nonumber \\
N \, \text{M5-branes} &: \quad \mathbb{R}_{t} \times M_{3} \times D \label{eqn:Kbrane} \\
|R| \, \text{M5'-branes} &: \quad \mathbb{R}_{t} \times L_{K} \times D \nonumber
\end{align}

Above, $M_{3}$ is a Seifert manifold, which will be assumed to be $M_{3}=S^{3}$ throughout this paper. $D \cong \mathbb{R}^{2} \subset M_{4} = TN$ is the ``cigar'' in the Taub-Nut space, $K$ represents a Wilson loop colored in a $SU(N)$ representation $R$, and $L_{K}$ is a Lagrangian submanifold of $T^{*}M_{3}$ intersecting $M_{3}$ along $K$. $M_{4}$ is twisted, so that $(z_{1},z_{2}) \rightarrow (q z_{1}, t z_{2})$ as one goes around the ``time'' direction. For general $M_{3}$, a prescription $q t = 1$ is necessary to preserve supersymmetry. When $M_{3}$ is a Seifert manifold, however, it has a semi-free $U(1)$ action on it, which defines a nowhere vanishing vector field on $M_{3}$. Then, at each point of $M_{3}$, we can define a 2-plane in the fiber of $T^{*}M$ such that the plane is orthogonal to the vector field. Then, the five-brane configuration of Equation \ref{eqn:Kbrane} has an extra rotation symmetry of these 2-planes, which will be denoted as $U(1)_{\beta}$ flavor symmetry. The extra $U(1)_{\beta}$ flavor symmetry can be used to preserve the supersymmetry, and now we can lift the $q t = 1$ prescription. 

\subsection{Construction of a vector field on $L_{\Gamma} \subset M_{3}$}
Now, let us check if the brane setup in Equation \ref{eqn:gamma_fivebrane} allows such refinement when $M_{3}$ is a Seifert manifold. We wish to find a $U(1)$ rotation symmetry in $T^{*}M_{3}$, which leaves $L_{\Gamma}$ invariant. The easiest way to obtain one is to construct a vector field on $M_{3}$ which is tangent to $\Gamma$. Then, the orthogonal vectors in the fiber of $T^{*}M_{3}$ comprise a rank-2 subbundle, whose restriction to $\Gamma \subset M_{3}$ is precisely $L_{\Gamma}$. We can inductively construct such a vector field by locally perturbing a pre-existing vector field around the junctions. 

\begin{figure}[htb]
\centering
\includegraphics{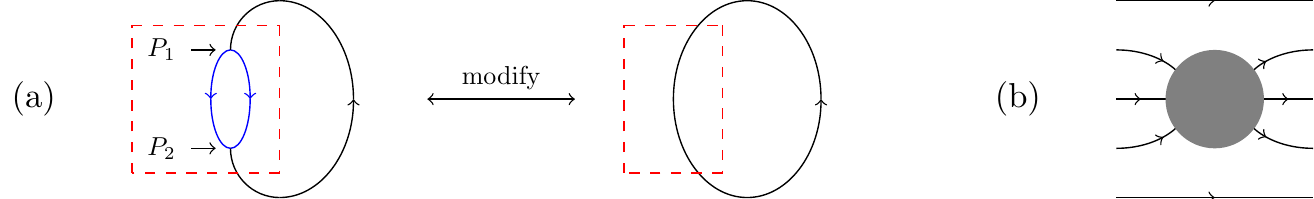}
\caption{(a) Two Wilson lines $W$ (left) and $\hat{W}$ (right) which are locally different, with and without two junctions in the red dashed square. (b) field line near the conductor in an external field.}
\label{fig:digon_mod}
\end{figure}

Let us start with a base case, a Wilson line $W$ with only two junctions (see Figure \ref{fig:digon_mod}(a).) As $W$ resembles an unknot outside the red dashed square (which is in fact a cylinder), we may consider another Wilson line $\hat{W}$ obtained from $W$ by replacing the interior of the red dashed square with a straight Wilson line. Since $\hat{W}$ is only an unknot, there is a nonvanishing vector field on $M_{3}$, which is tangent to $\hat{W}$. We can safely assume that this ``external'' vector field is uniform inside the red dashed square, pointing downwards in Figure \ref{fig:digon_mod}(a). To define a vector field tangent to $\Gamma$ inside the red dashed square, we consider the vector field as ``electric'' and place two small spherical conductors $P_{1}$ and $P_{2}$ at each junctions (see Figure \ref{fig:digon_mod}(b).) Upon the insertion, the field lines are normal to the spherical shell of conductors and vanish nowhere except for the interior of $P_{1}$ and $P_{2}$. Now, draw radial field lines from the position of junctions towards the boundary of $P_{1}$ and $P_{2}$. The resultant vector field vanishes only at the junctions, is tangent to the trivalent graph, and differs only locally from that of $\hat{W}$. 

\begin{figure} [htb]
\centering
\includegraphics{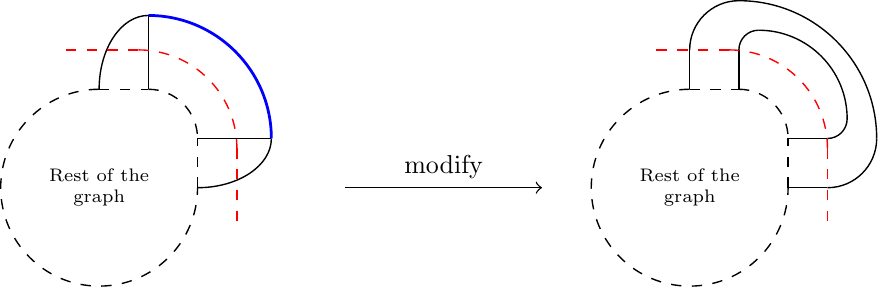}
\caption{Induction step from $2k$ junctions to $2k+2$ junctions.} 
\label{fig:2k_junction}
\end{figure}

Next, let us assume that we can define a vector field tangent to any graphs with $2k$ junctions (junctions must always appear in even numbers, for all of them are trivalent.) Arbitrarily chose a trivalent graph with $2k+2$ junctions. Pick any two adjacent two junctions, and get rid of them as in Figure \ref{fig:2k_junction}.

Here, we are considering the two adjacent junctions to be ``close'', in a sense that they are contained in a small 3-ball. In the modified graph, the new edges are drawn so that they are parallel to the original (thick blue) edge, in a sense that they do not introduce any unnecessary crossings. The resultant graph has only $2k$ junctions, so by assumption there is a vector field with all the desired properties. Above the red line, in particular, the vector field will be parllel and tangent to the two newly added edges, for the two edges are ``close enough'' to the original blue edge. Therefore, as we insert two zero-size conducting spheres at the junctions, we obtain the desired vector field for the original trivalent graph. 

\subsection{Kinematics of refined junctions}
\label{subsec:kinematics}

Before proceeding further, let us briefly discuss the kinematics of refined junctions. So far, the representations of Wilson lines were not specified. In ordinary Chern-Simons theory, the representations $R$ are the integrable representations of $SU(N)_{k}$, and the dimension of the space of conformal blocks associated to a 3-puncutred $S^{2}$ coincides with the Littlewood-Richardson coefficient $N_{ijk}$. In the refined case, although $N_{ijk}$ is no more an integer, the Hilbert space remains the same (thus, $N_{ijk}$ is no more considered as the dimension of the Hilbert space.) In particular, the Hilbert space associated to a solid torus $H_{T^{2}}$ remains unchanged upon refinement, which is spanned by vectors $|R \rangle$ obtained by inserting Wilson line in an integrable representation $R$ inside the solid torus. We can furthermore endow the Hilbert space with an inner product, where the metric $\langle R | R' \rangle = g_{R}^{R'}$ is diagonal and moreover Hermitian. 

If we put two Wilson lines colored in $R$ and $R'$ inside a solid torus, a path integral fixes a vecor in $H_{T^{2}}$. Next, consider another solid torus without any Wilson lines inside. Gluing the two solid tori via the orientation reversing identity morphism, we obtain $S^{2} \times S^{1}$ with two Wilson lines in $R$ and $R'$ along the non-contractible 1-cycle. The partition function is given by:
$$\langle 0 | R,R' \rangle = Z(S^{2} \times S^{1} ; R,R') = \langle R^{*} | R' \rangle = g_{R}^{R'}.$$
In other words, $| R,R' \rangle$ can be written as $g_{R}^{R'}/g_{0}^{0}|0 \rangle + \cdots$. Since $g_{R}^{R'}$ is diagonal, we can see that $|R,R' \rangle$ has a nontrivial $|0 \rangle$ component if and only if $R^{*} = R'$. In other words, the dimension of $H_{\{ S^{2};R,R' \}}$ is nonzero if and only if $0 \in R \otimes R'$. We can do the same with three Wilson lines to get $|R,R',R'' \rangle = N_{R,R',R''}/g_{0}^{0}|0\rangle + \cdots$ and conclude that the dimension of $H_{\{ S^{2};R,R',R'' \}}$ is nonzero if and only if $0 \in R \otimes R' \otimes R''$. 

Therefore, the ``charge conservation'' for refined Chern-Simons theory works exactly the same way as in ordinary Chern-Simons theory, and we can furthermore argue that the dimension of $H_{\{ S^{2};R,R',R'' \}}$ is equal to the dimension of $Inv_{G}(R \otimes R' \otimes R'')$. In other words, the three Wilson lines in representations $R,R',R''$ are allowed to form a junction only when $0 \in R \otimes R' \otimes R''$. This ``charge conservation'' argument extends to arbitrary number of punctures, and we conclude that the same set of refined Wilson lines on a punctured sphere will satisfy a linear relation as in the ordinary Chern-Simons theory, while the coefficient may be refined. In particular, all dynamics of junctions involving the dual representations are identical. For instance, a Wilson line colored in $R$ is equivalent to the Wilson line colored in $\bar{R}$ with reverse orientation. 

\begin{figure} [htb]
\centering
\includegraphics{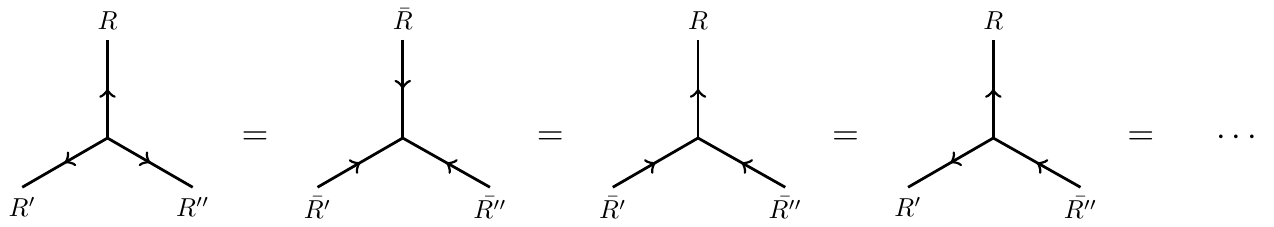}
\caption{Wilson lines representing the same vector in $H_{S^{2};R,R',R''}$, where $0 \in R \otimes R' \otimes R''$.}
\label{fig:dual}
\end{figure}

We assume throughout this paper that we can normalize the junction degrees of freedom so that the Wilson lines in Figure \ref{fig:dual} represent the same vector in $H_{\{ S^{2};R,R',R'' \}}$. This is only an assumption, for we do not fully understand the junction degrees of freedom in refined Chern-Simons theory. Unlike ordinary Chern-Simons theory, the refined theory does not have a gauge theory description, so we cannot simply place gauge invariant tensors at junctions to define a gauge invariant observable. But then, how do we know whether there are junction degrees of freedom to normalize? In \cite{CGR}, the gauge invariant tensors were normalized so that junctions of Wilson lines $\Gamma$ categorify to the interface of Landau-Ginzburg models on $\mathbb{R}_{t} \times \Gamma$. For exampe, the ``digon removal'' relation in Figure \ref{fig:relsPlanar} is categorified to a fusion of two interfaces in Landau-Ginzburg models. Chiral superfields and boundary BRST operators of the B-type defects constitute the matrix factorization descriptions of the interfaces, and the fusion of interfaces corresponds to a tensor product of matrix factorizations. In case of the fusion shown in the digon removal relation, its matrix factorization decomposes into a direct sum of degree-shifted identity defects, whose graded dimension precisely equals the proportionality factor on the RHS. Since the $U(1)_{\beta}$ flavor symmetry is present in the five-brane setup of \cite{CGR}, the matrix factorization of the interfaces will be refined as well, and we can interpret the normalization of junctions as the associated graded dimension. Yet, the refined theory of surface operators is a distant goal, and we leave it for future works.

\section{Global corrections: $\hat{N}_{RR'}^{Q}$ and $\gamma_{RR'}^{Q}$}
\label{sec:global}
We will shortly discuss local relations which comprise the $\beta$-deformed $\dot{U}_{q}(\mathfrak{sl}_{m})$ relations. But before proceeding further, we have to discuss certain ``global corrections'', which is an exquisite feature of the refined theory. These global corrections arise when homotopy of Wilson lines wrap the entire orbit of $U(1)$ action on $M_{3}$. Let us illustrate two simple cases.

Consider two parallel Wilson lines in representation $R$ and $R'$ inside a closed 3-ball. The vector field is uniformly upward, but locally perturbed in the vicinity of junctions. Dimension of the Hilbert space is equal to $\sum_{Q\in R \otimes R'} N_{RR'}^{Q}$, summed over irreducible representations $Q$. When all $Q$'s are non-degenerate (that is, $N_{RR'}^{Q} = 1$), the Wilson lines in the LHS of Figure \ref{fig:idempotent} satify a local relation which can be considered as an idempotent decomposition of the identity morphism $id: (R \otimes R') \rightarrow (R \otimes R')$:

\begin{figure}[htb]
\centering
\includegraphics{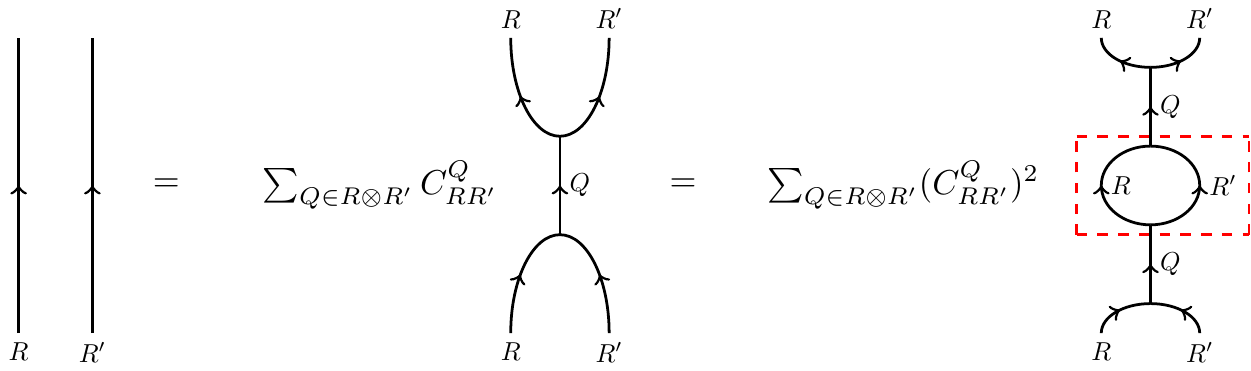}
\caption{Idempotent decomposition of identity Wilson lines.}
\label{fig:idempotent}
\end{figure}

\begin{figure} [htb]
\centering
\includegraphics{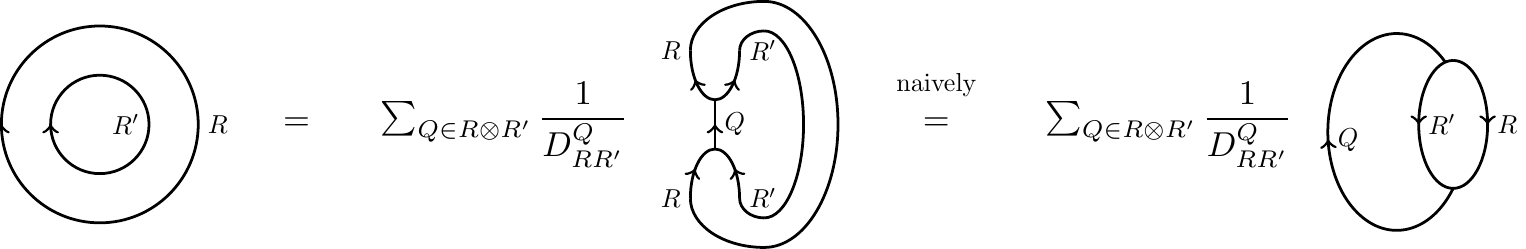}
\caption{A ``naive'' equality involving global homotopy which wraps around an entire orbit of $U(1)$ action on $S^{3}$.}
\label{fig:naive_equality}
\end{figure}

The second identity is obtained by vertically composing the first identity, and the summation $\sum_{Q} \sum_{Q'}$ simplifies to $\sum_{Q}$ by the fact that $H_{\{ S^{2},Q,\bar{Q'} \}}$ is non-vanishing if and only if $Q=Q'$. Moreover, since  $H_{\{ S^{2},Q,\bar{Q} \}}$ is one-dimensional, the Wilson line configuration with a ``digon'' in the red dashed rectangle in Figure \ref{fig:idempotent} must be proportional to a straignt Wilson line colored in $Q$. Let us put the proportionality constant as $D_{RR'}^{Q}$. Then, the following equality must hold: 
$$C_{RR'}^{Q} = (C_{RR'}^{Q})^{2}D_{RR'}^{Q} \quad \Rightarrow \quad C_{RR'}^{Q} = \dfrac{1}{D_{RR'}^{Q}}.$$
However, we will encounter a contradiction if we allow the ``digon removal'' relation to hold globally. Let us close the Wilson lines of the first identity in Figure \ref{fig:idempotent}, as shown in Figure \ref{fig:naive_equality}. Upon gluing the two 3-balls, the LHS becomes two disjoint unknots colored in $R$ and $R'$, while the RHS becomes a linear sum of Wilson lines with an extended digon. If we smoothly deform the extended digon to the other side, we obtain the second equality in Figure \ref{fig:naive_equality}.

But as we apply the digon removal relation on the RHS of Figure \ref{fig:naive_equality}, we obtain the following identity.
\begin{equation}
M_{R}(q_{2}^{\rho_{N}};q_{1},q_{2})M_{R'}(q_{2}^{\rho_{N}};q_{1},q_{2}) = \sum_{Q \in R \otimes R'} M_{Q}(q_{2}^{\rho_{N}};q_{1},q_{2}),
\label{eqn:false_decom}
\end{equation}
where $M_{R}(q_{2}^{\rho_{N}};q_{1},q_{2})$ is a Macdonald polynomial for representation $R$ of SU(N), with $\rho_{N} = \big( (N-1)/2, (N-3)/2, \cdots, (1-N)/2 \big)$. The two parameters $q_{1}$ and $q_{2}$ are $\beta$-deformation of $q = e^{2 \pi i / (k + N)}$ in orindary Chern-Simons theory, $q_{1} = e^{2 \pi i / (k + \beta N)}, q_{2} = e^{2 \pi i \beta / (k + \beta N)}.$ However, this identity is not true in general! The reason for the obvious contradiction is very simple: the LHS of the identity implies that the two unknots are disjoint. In other words, we can move them far apart, and evaluate their expectation values using the connected sum formula of \cite{Witten89}. However, the global homotopy of shrinking the digon goes around entire $U(1)$ orbit of $M_{3}$, and it forces the two unknots to be close to each other (for digons were local, \textit{i.e.}, close to each other.) The two interpretations are obviously different, and we can remedy the obvious contradiction by implementing a global correction which relates the two ``globally'' homotopic Wilson lines:

\begin{figure} [htb]
\centering
\includegraphics{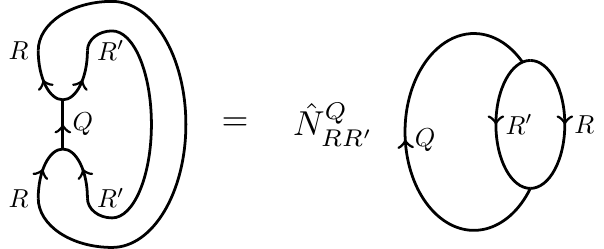}
\caption{``Global correction'' from a homotopy which wraps around a $U(1)$ orbit of $S^{3}$.}
\label{fig:global_correction}
\end{figure}
The generalized Littlewood-Richardson coefficient $\hat{N}_{RR'}^{Q}$ is not equal to $1$ in general \cite{Mac}, and the global correction changes Equation \ref{eqn:false_decom} into:

\begin{equation}
M_{R}(q_{2}^{\rho_{N}};q_{1},q_{2})M_{R'}(q_{2}^{\rho_{N}};q_{1},q_{2}) = \sum_{Q \in R \otimes R'} \hat{N}_{RR'}^{Q} M_{Q}(q_{2}^{\rho_{N}};q_{1},q_{2}).
\label{eqn:true_decom}
\end{equation}
And Equation \ref{eqn:true_decom} holds for any $R$ and $R'$. Again, the extra factor of $\hat{N}_{RR'}^{Q}$ seemingly destroys the topological nature of the theory, but in fact the expectation values of two Wilson lines of Figure \ref{fig:global_correction} are evaluated in two different ways: on LHS we consider the two junctions as local perturbations of two parallel Wilson lines colored in $R$ and $R'$, while on RHS they are considered as local perturbations of a single Wilson line colored in $Q$. In terms of surface operators and their interfaces, the LHS evaluates the trace of $(R \otimes R' \rightarrow Q) * (Q \rightarrow R \otimes R')$ while the RHS evaluates the trace of $(Q \rightarrow R \otimes R') * (R \otimes R' \rightarrow Q)$ (here, the trace means the graded dimension of identity morphisms in the direct sum decomposition.) The naive equality asserts that the two must be equal, while the Equation \ref{eqn:true_decom} shows that a non-trivial correction is inevitable.

\begin{figure} [htb]
\centering
\includegraphics{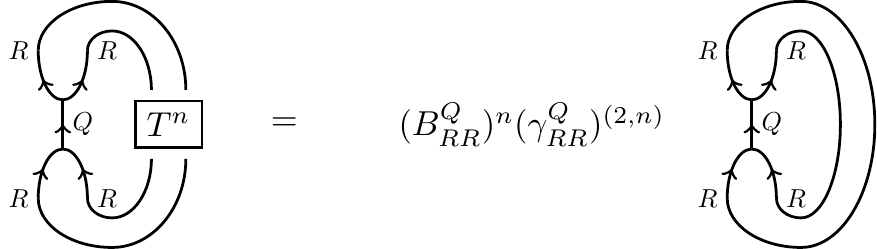}
\caption{Gamma factor as a global correction of braided idempotents}
\label{fig:gamma}
\end{figure}

The gamma factor which appears in the refined expectation value of torus knots \cite{BMMSS} also acquires an interpretation this way: it can be considred as a ``global correction'' to the braiding operator near the vertex of $R,R,Q$, as shown in Figure \ref{fig:gamma}.
In Figure \ref{fig:gamma}, $T^{n}$ represents a righthanded $(2,n)$-braid. Such a global correction is necessary to prevent another naive equality which fails in general. Unfortunately, we cannot derive the global corrections in the refined theory, as it requires a complete understanding of junction degrees of freedom.

\section{Refined skew-Howe duality and $\beta$-deformed $\dot{U}_{q}(\mathfrak{sl}_{m})$}
\begin{figure} [htb]
\centering
(associativity) \quad \raisebox{-0.5\height}{\includegraphics{CS_associativity.pdf}} \\[1.5ex]
(digon removal) \quad  \raisebox{-0.5\height}{\includegraphics{CS_digon.pdf}} \\[1.5ex]
($[E,F]$ relation) \quad \raisebox{-0.5\height}{\includegraphics{EF.pdf}} \\[2.5ex]
(braiding relations) \\[1.5ex]
\includegraphics{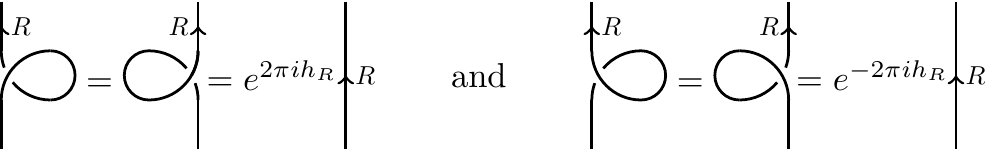} \\[1.5ex]
\includegraphics{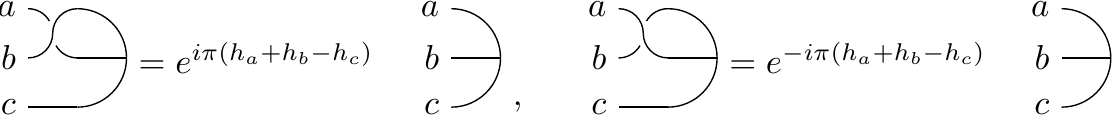} \\[1.5ex]
\includegraphics{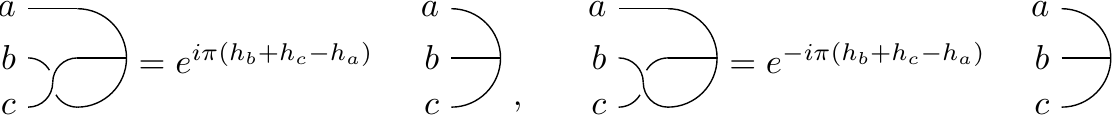}
\caption{Local relations of Wilson lines which determine the expectation values of all trivalent graphs}
\label{fig:rels}
\end{figure}

In this section, we propose refined web relations of Wilson lines. It was shown in \cite{CGR} that the entire list of web relations in ordinary Chern-Simons theory can be derived from the relations of Figure \ref{fig:rels}.

We will refine the commutation relation shortly, so let us focus on a refinement of the other relations in Figure \ref{fig:rels}. As was argued in subection \ref{subsec:kinematics}, these relations are refined by a $\beta$-deformation of their coefficients. We provide here a proposal (there might as well be other choices) to refine the relations in Figure \ref{fig:rels}, following the dictionary (3.27) of \cite{FGS}. 

\begin{itemize}
\item associativity relation: in ordinary Chern-Simons theory, the Wilson lines in the associativity relation are categorified to the fusion of interfaces, whose behaviors are analogous to partial symmetrization of $i+j+k$ variables. In this viewpoint, the associativity relation implies the order of partial symmetrization is not important. We may expect the same to hold in the refined matrix factorizations, and we leave the associativity relation unaltered upon refinement.
\item digon removal: the key intuition is that when $i+j = N$, the digon removal relation coincides with the circle removal relation. Therefore, the natural choice will be replacing the quantum binomial $\quantumbinomial{i+j}{i}$ by $M_{\wedge^{i}\square}(q_{2}^{\rho_{i+j}};q_{1},q_{2})$, where $\rho_{i+j}$ is given by $\Big(\tfrac{(i+j)-1}{2}, \tfrac{(i+j)-3}{2}, \cdots, \tfrac{1-(i+j)}{2} \Big)$. Here, we have treated $q_{1}$ and $q_{2}$ as formal variables.
\item braiding relations: the phase factors $e^{i \pi h_{R} } = q^{C_{2}(R)}$ allows a natural way to upgrade via (3.27) of \cite{FGS}, namely, $q_{1}^{\tfrac{1}{2}||R||^{2}}q_{2}^{-\tfrac{1}{2}||R^{t}||^{2}}q_{2}^{\tfrac{N}{2}|R|}q_{1}^{-\tfrac{1}{2N}|R|^{2}}$. Since the Wilson lines of interest are all vertically framed, we must not multiply the braid operator by an extra factor (as in \cite{FGS}) to canonically frame the Wilson lines. 
\end{itemize}

Now we can proceed to derive other $\beta$-deformed quantum group relations. To avoid clutter, we will use the following short-hand notations:
\begin{gather*}
M_{k}^{m} := M_{\wedge^{k}\square}(q_{2}^{\rho_{m}};q_{1},q_{2}) = \prod_{i=1}^{k} \dfrac{q_{2}^{\tfrac{m-i+1}{2}} - q_{2}^{-\tfrac{m-i+1}{2}}}{q_{2}^{\tfrac{k-i+1}{2}} - q_{2}^{-\tfrac{k-i+1}{2}}}, \\
[j]_{q_{1}} = \dfrac{q_{1}^{j/2}-q_{1}^{-j/2}}{q_{1}^{1/2}-q_{1}^{-1/2}}, \quad [j]_{q_{2}} = \dfrac{q_{2}^{j/2}-q_{2}^{-j/2}}{q_{2}^{1/2}-q_{2}^{-1/2}}, \\[1.5ex]
[j]_{\wedge} = \dfrac{q_{1}^{1/2}q_{2}^{(j-1)/2}-q_{1}^{-1/2}q_{2}^{-(j-1)/2}}{q_{1}^{1/2}-q_{1}^{-1/2}}, \quad [j]_{s} = \dfrac{q_{2}^{1/2}q_{1}^{(j-1)/2}-q_{2}^{-1/2}q_{1}^{-(j-1)/2}}{q_{2}^{1/2}-q_{2}^{-1/2}}
\end{gather*}
which allows us to write some generalized Littewood-Richardson coefficients in a simpler form: $\hat{N}_{\square,\wedge^{j}\square}^{\wedge^{(j+1)}\square} = [j+1]_{q_{2}}/[j+1]_{\wedge}$. Notice that $M_{k}^{m}$ reduces to $\quantumbinomial{m}{k}$ in the unrefined limit $q_{1} = q_{2}$. Also, $[j]_{\wedge}$ and $[j]_{s}$ are related via $q_{1} \leftrightarrow q_{2}$, which will later be seen as an analogue of the level-rank duality in the ordinary Chern-Simons theory.

\subsection{Refined commutation and Serre relations}
\label{sec:anti}
\begin{figure} [htb]
\centering
\includegraphics{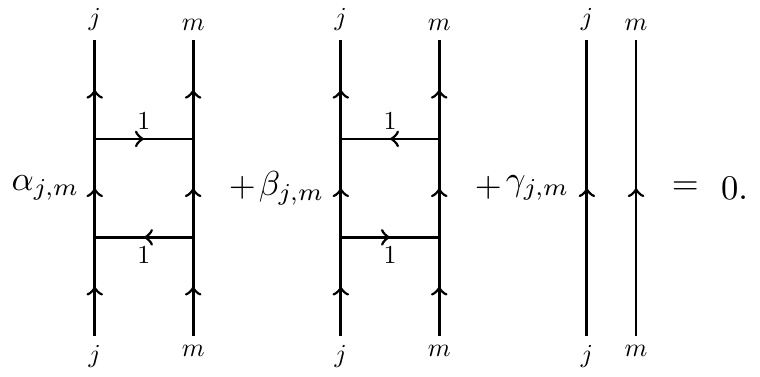}
\caption{A linear relation satisfied by three Wilson line configurations, which are colored in antisymmetric representations.}
\label{fig:EF_undet}
\end{figure}
\begin{figure} [htb]
\centering
\includegraphics{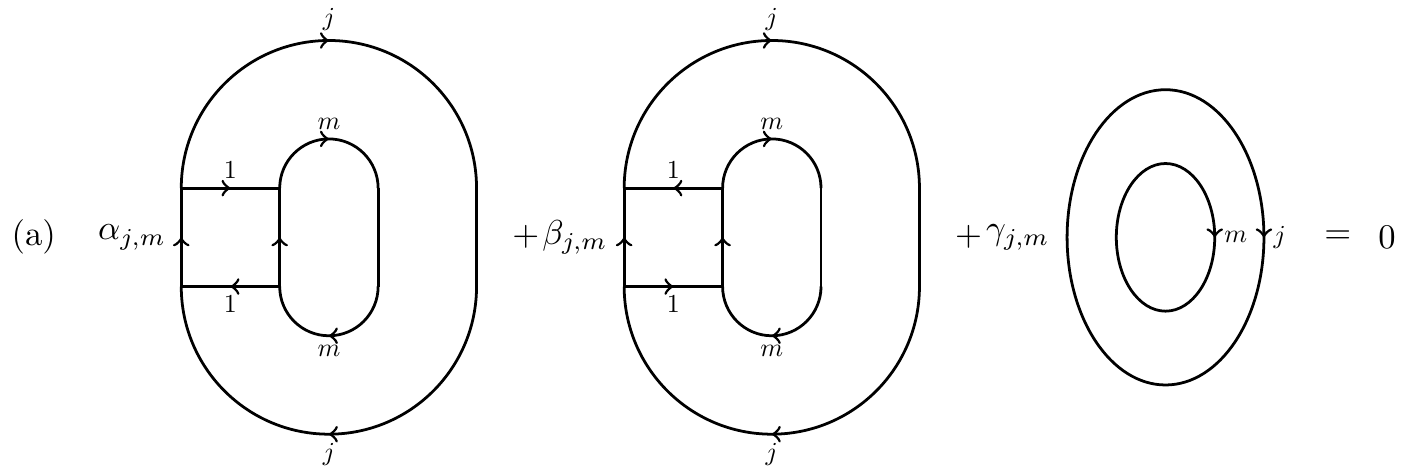}
\includegraphics{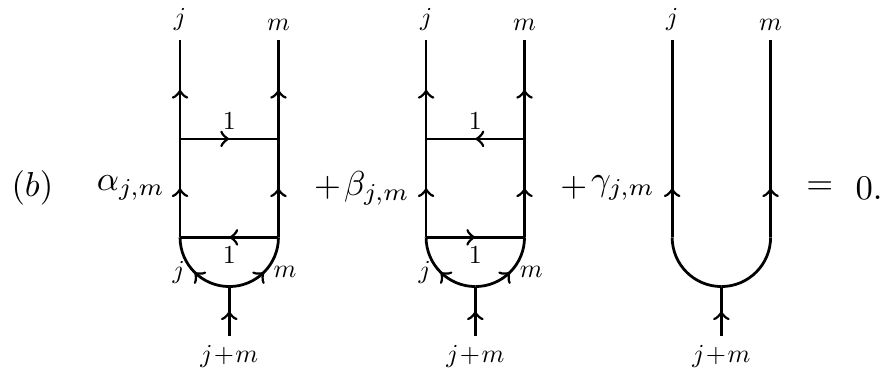}
\caption{(a) Closing Wilson lines along orbits of $U(1)$ action on $S^{3}$. (b) Attaching a Wilson line $\wedge^{j+m}\square \rightarrow \wedge^{j}\square \otimes \wedge^{m}\square$ from below.}
\label{fig:EF_close}
\end{figure}

Now, we study the refinement of the $[E,F]$ relation. Just as in the ordinary Chern-Simons theory, the three Wilson lines in Figure \ref{fig:EF_undet} satisfy a linear relation (here, $j,m$ stand for $\wedge^{j}\square$ and $\wedge^{m}\square$, resp.) Next, we glue them with two other Wilson line configurations, as shown in Figure \ref{fig:EF_close}.
In Figure \ref{fig:EF_close}(a), we have taken the ``trace'' by gluing Wilson lines inside $S^{3} \setminus B^{3}$ so that the $j$- and $m$-colored Wilson lines wrap around orbits of $U(1)$ action on $S^{3}$. In Figure \ref{fig:EF_close}(b), we have simply attached another 3-ball which contains a junction of Wilson line, $\wedge^{j+m}\square \rightarrow \wedge^{j}\square \otimes \wedge^{m}\square$. 

\begin{figure}[htb]
\centering
\includegraphics{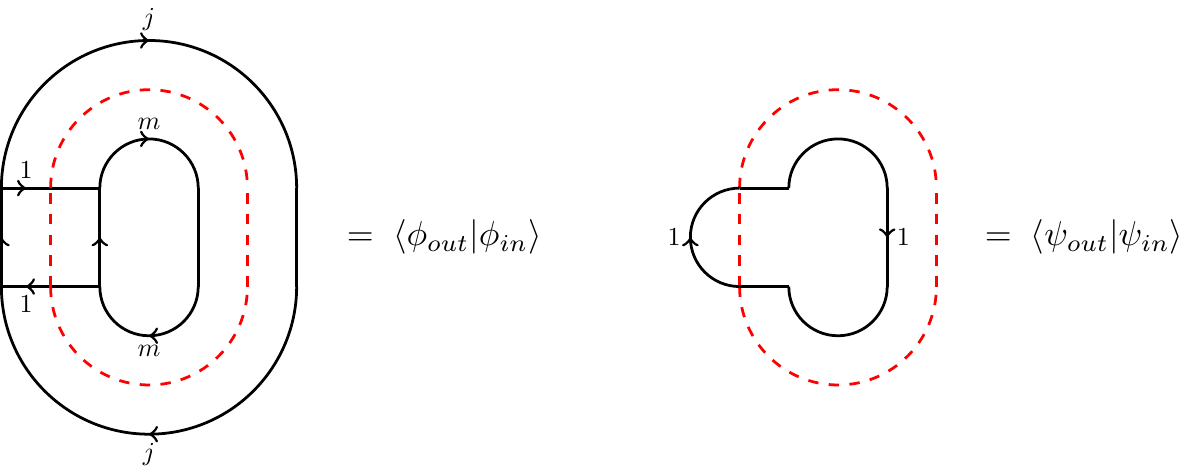} \\[1.5ex]
\includegraphics{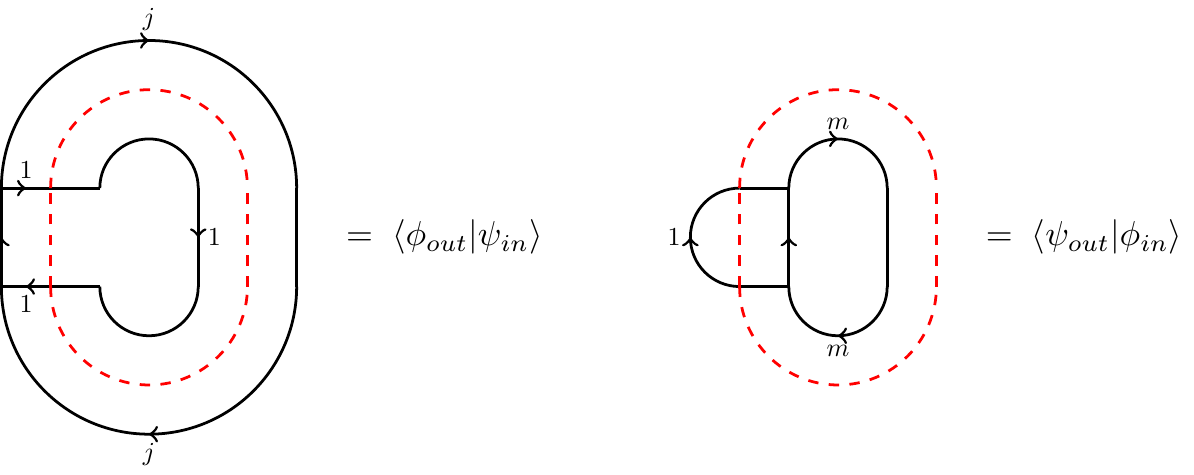} \\[3ex]
$\langle \phi_{out} | \phi_{in} \rangle =  \dfrac{\langle \phi_{out} | \psi_{in} \rangle \langle \psi_{out} | \phi_{in} \rangle}{\langle \psi_{out} | \psi_{in} \rangle} \quad \text{(connected sum formula)}$
\caption{Evaluation of the $\alpha_{j,m}$-term via connected sum formula.}
\label{fig:connected}
\end{figure}

Using associativity and digon removal relations in Figure \ref{fig:EF_close}(b), one can immediately obtain a linear relation of coefficients $\alpha_{j,m}, \beta_{j,m}, \gamma_{j,m}$. We can obtaion another from Figure \ref{fig:EF_close}(a) as follows: let us first consider the $\alpha_{j,m}$-term in Figure \ref{fig:EF_close}(a). Consider a 3-ball (depicted as a red dashed circle in the top left corner of Figure \ref{fig:connected}), which contains $m$- and $(m-1)$- colored Wilson lines. A path integral on the closed 3-ball determines a vector $|\phi_{in} \rangle \in H_{\{ S^{2};\square,\bar{\square} \} }$. Performing a path integral on the complement determines a vector in the dual Hilbert space, $\langle \phi_{out} | \in H_{\{ S^{2};\square,\bar{\square} \} }^{*}$. Then, the refined expectation value of the $\alpha_{j,m}$-term is simply their inner product, $\langle \phi_{out} | \phi_{in} \rangle$ (see top left corner of Figure \ref{fig:connected}.) Now, consider a separate configuration, in which a Wilson loop colored in $\square$ wraps an entire $U(1)$ orbit of $S^{3}$. Cut the 3-sphere into two 3-balls so that their boundaries have two punctures (top right corner of Figure \ref{fig:connected}.) Then, the interior of the red dashed circle will determine a vector $| \psi_{in} \rangle \in H_{\{ S^{2};\square,\bar{\square} \}}$, while the exterior determines a vector $\langle \psi_{out} | \in H_{\{ S^{2};\square,\bar{\square} \}}^{*}$. Now that $H_{\{ S^{2};\square, \bar{\square} \}}$ is one-dimensional, $|\phi_{out} \rangle \propto | \psi_{out} \rangle$ and $\langle \phi_{in} | \propto \langle \psi_{in} |$. Therefore, we can write the refined expectation value of the $\alpha_{j,m}$-term $\langle \phi_{out} | \phi_{in} \rangle$, in terms of $\langle \phi_{out} | \psi_{in} \rangle$,  $\langle \psi_{out} | \phi_{in} \rangle$ and $\langle \psi_{out} | \psi_{in} \rangle$ as shown in the connected sum formula of Figure \ref{fig:connected}. We can compute the refined expectation values $\langle \phi_{out} | \psi_{in} \rangle$,  $\langle \psi_{out} | \phi_{in} \rangle$ and $\langle \psi_{out} | \psi_{in} \rangle$ by digon removal relations along with global corrections when needed.   

Doing the same with $\beta_{j,m}$-term, we obtain two linear relations for three undetermined coefficients $\alpha_{j,m}, \beta_{j,m}, \gamma_{j,m}$, from which we can determine their ratio:
\begin{gather}
\alpha_{j,m} N_{1,j}^{j+1}M_{1}^{j+1}M_{j+1}^{N}M_{1}^{m}M_{m}^{N} + \beta_{j,m} N_{1,m}^{m+1}M_{1}^{j}M_{j}^{N}M_{1}^{m+1}M_{m+1}^{N} + \gamma_{j,m} M_{1}^{N}M_{j}^{N}M_{m}^{N} = 0. \\
\alpha_{j,m} M_{1}^{j+1}M_{1}^{m} + \beta_{j,m} M_{1}^{j}M_{1}^{m+1} + \gamma_{j,m} = 0. \\[1.5ex]
\Rightarrow \alpha_{j,m} : \beta_{j,m} : \gamma_{j,m} = -\frac{[j+1]_{\wedge}}{[j+1]_{q_{2}}} : \frac{[m+1]_{\wedge}}{[m+1]_{q_{2}}} : [m-j]_{q_{2}}.
\label{eqn:antiEF}
\end{gather}
Note that in the limit $\beta \rightarrow 1$, the ratio becomes $-1:1:[m-j]_{q}$, which reproduces the commutation relation in ordinary Chern-Simons theory.

We wish to provide another example, the refined Serre relation. By repeatedly applying refined commutation relations, we can derive the refined square switch relation (Figure \ref{fig:square_switch}), where coefficients $A$ and $B$ in Figure \ref{fig:square_switch} are given by:
\begin{figure} [htb]
\centering
\includegraphics{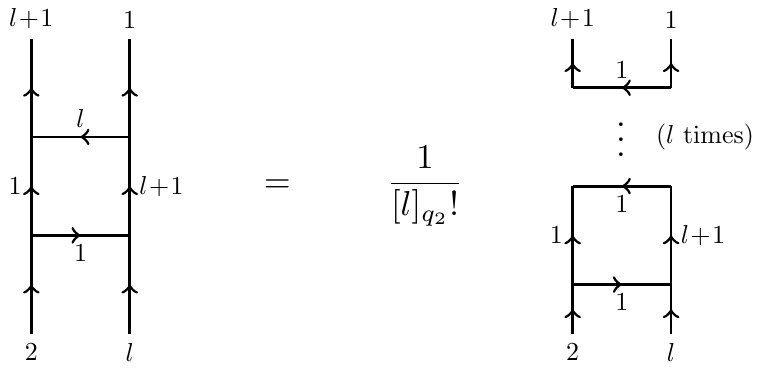} \\[1.5ex]
\includegraphics{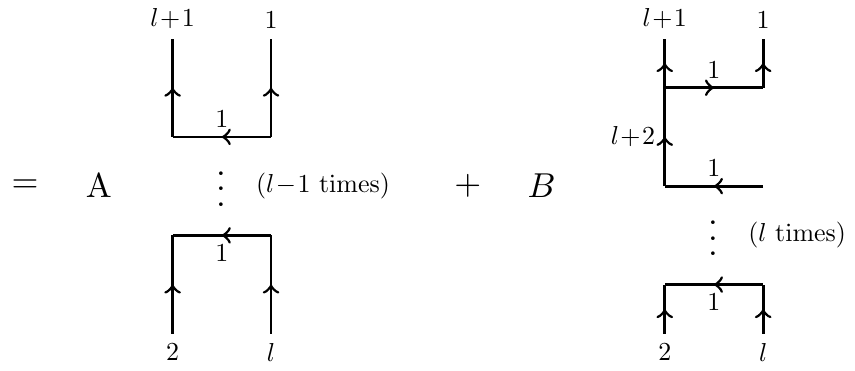}
\caption{A square switch relation from which the Serre relation follows.}
\label{fig:square_switch}
\end{figure}

\begin{gather*}
A = \dfrac{1}{[l]_{q_{2}}!}\bigg[ (-\frac{\gamma_{2,l}}{\beta_{2,l}}) + (-\frac{\alpha_{2,l}}{\beta_{2,l}})(-\frac{\gamma_{3,l-1}}{\beta_{3,l-1}}) + \cdots +  (-\frac{\alpha_{2,l}}{\beta_{2,l}})\cdots (-\frac{\alpha_{l,2}}{\beta_{l,2}})(-\frac{\gamma_{l+1,1}}{\beta_{l+1,1}}) \bigg], \\
= \dfrac{1}{[l]_{q_{2}}!}(-\frac{\gamma_{l+1,1}}{\beta_{l+1,1}}) = \dfrac{1}{[l]_{q_{2}}!}\frac{[l]_{q_{2}}[2]_{q_{2}}}{[2]_{\wedge}},  \quad \text{and} \\
B = \dfrac{1}{[l]_{q_{2}}!}  (-\frac{\alpha_{2,l}}{\beta_{2,l}}) (-\frac{\alpha_{3,l-1}}{\beta_{3,l-1}}) \cdots (-\frac{\alpha_{l+1,1}}{\beta_{l+1,1}}) = \dfrac{1}{[l]_{q_{2}}!}(-\frac{\alpha_{l+1,1}}{\beta_{l+1,1}}) = \dfrac{1}{[l]_{q_{2}}!}\frac{[l+2]_{\wedge}[2]_{q_{2}}}{[l+2]_{q_{2}}[2]_{\wedge}}.
\end{gather*}
Then, it remains to use associativity and digon removal relations to turn $(l-1)$ and $l$ parallel $\square$-colored Wilson lines into a single $\wedge^{l-1}\square$- and $\wedge^{l}\square$-colored Wilson lines. Applying the resultant ``sqaure switch'' relation as in \cite{CKM}, we obtain the refined Serre relation (Figure \ref{fig:Serre}):

\begin{figure} [htb]
\centering
\includegraphics{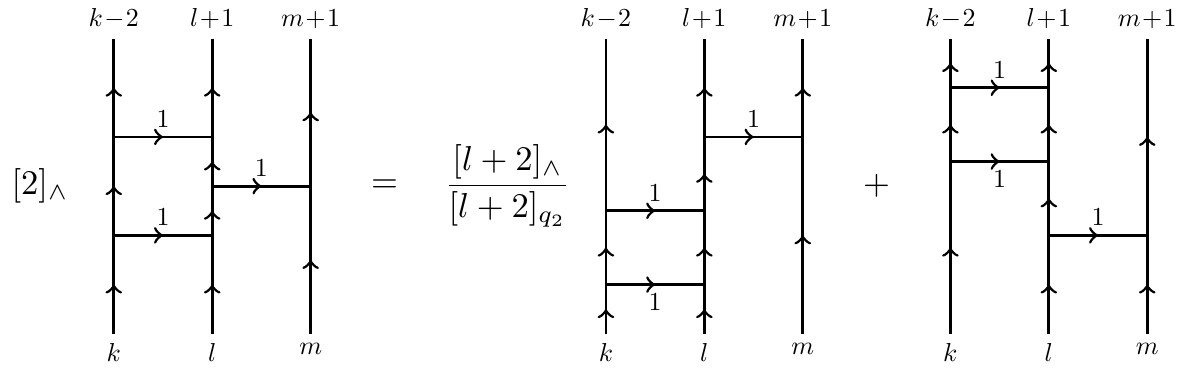}
\caption{The refined Serre relation.}
\label{fig:Serre}
\end{figure}

\subsection{Rescaling of quantum group generators}
Under the skew-Howe duality functor \cite{CKM, CGR}, the commutation relation in ordinary Chern-Simons theory can be recast as: 
$$[E_{i},F_{i}]1_{\lambda} = [k_{i+1}-k_{i}]_{q}1_{\lambda},$$
where $\lambda = (k_{1}, \cdots, k_{n})$ is a weight vector of $\mathfrak{sl}_{N}$. 

\begin{figure} [htb]
\centering
\includegraphics{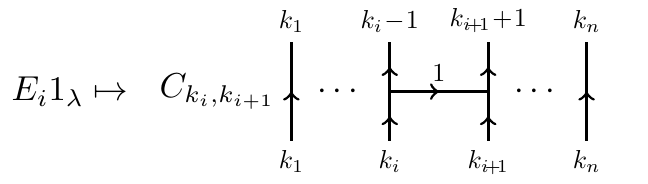}
\includegraphics{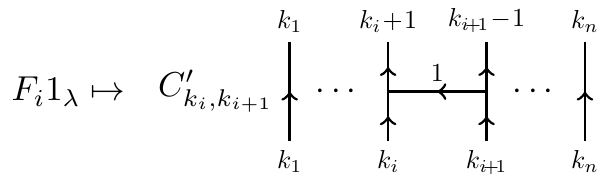}
\caption{Refined skew-Howe duality functor for $\lambda = (k_{1}, \cdots, k_{n})$.}
\label{fig:rescale}
\end{figure}

In the refined case, however, one immediately sees that the refined commutation relation (Figure \ref{fig:EF_undet} and Equation \ref{eqn:antiEF}) is not a commutation relation but a ``twisted'' commutation relation:
\begin{equation}
\frac{[k_{i}+1]_{\wedge}}{[k_{i}+1]_{q_{2}}}E_{i}F_{i}1_{\lambda} - \frac{[k_{i+1}+1]_{\wedge}}{[k_{i+1}+1]_{q_{2}}}F_{i}E_{i}1_{\lambda} = [k_{i+1} - k_{i}]_{q_{2}}1_{\lambda}.
\label{eqn:twistedEF}
\end{equation}
We can interpret Equation \ref{eqn:twistedEF} in two ways: (1) the $\beta$-deformation of $\dot{U}_{q}(\mathfrak{sl}_{m})$ involves twisted commutators instead of the ordinary commutators, or (2) the skew-Howe duality functor should also be refined so that the refined $[E,F]$ relation becomes a commutation relation. In this section, we propose a refinement of skew-Howe duality functor which will turn the refined $[E,F]$ relation into a commutation relation. Let us ``rescale'' the quantum group generators up to a proportionality constant as in Figure \ref{fig:rescale}.

Upon such a rescaling, Equation \ref{eqn:twistedEF} will read as follows:
\begin{equation}
\frac{[k_{i}+1]_{\wedge}}{[k_{i}+1]_{q_{2}}}\frac{1}{C_{k_{i}+1,k_{i+1}-1}C'_{k_{i},k_{i+1}}}E_{i}F_{i}1_{\lambda} - \frac{[k_{i+1}+1]_{\wedge}}{[k_{i+1}+1]_{q_{2}}}\frac{1}{C'_{k_{i}-1,k_{i+1}+1}C_{k_{i},k_{i+1}}}F_{i}E_{i}1_{\lambda} = [k_{i+1} - k_{i}]_{q_{2}}1_{\lambda}.
\label{eqn:rescaledEF}
\end{equation}
To get an ordinary commutation relation, the coefficient of $EF$-term and $FE$-term must be equal with an opposite sign:
$$\frac{C_{k_{i},k_{i+1}} C'_{k_{i}-1,k_{i+1}+1}}{C_{k_{i}+1,k_{i+1}-1} C'_{k_{i},k_{i+1}}} = \frac{[k_{i+1}+1]_{\wedge}[k_{i}+1]_{q_{2}}}{[k_{i+1}+1]_{q_{2}}[k_{i}+1]_{\wedge}}.$$
Notice that the denominator and nominator of LHS are related by a simple exchange of $k_{i} \leftrightarrow k_{i}+1$ and $k_{i+1} \leftrightarrow k_{i+1}-1$. The following choice of ``rescaling'' puts the refined $[E,F]$ relation into an ordinary commutatin relation:
\begin{gather}
C_{k_{i},k_{i+1}} = [k_{i}]_{\wedge}!/[k_{i+1}+1]_{q_{2}}!, \quad C'_{k_{i},k_{i+1}} = [k_{i+1}]_{\wedge}!/[k_{i}+1]_{q_{2}}!\\[1.5ex]
\Rightarrow \quad [E_{i},F_{i}]1_{\lambda} = [k_{i+1}-k_{i}]_{q_{2}}\frac{[k_{i}]_{\wedge}![k_{i+1}]_{\wedge}!}{[k_{i}]_{q_{2}}![k_{i+1}]_{q_{2}}!}1_{\lambda}.
\label{eqn:omg}
\end{gather}

Correction on the RHS is ominous from the categorification perspective, for it is a huge rational function of $q_{1}$ and $q_{2}$ which is not readily associated to a graded dimension. One may wish for a better way to rescale the quantum group generators, but it is shown in Appendix \ref{appen:digon} that there is an inevitable trade-off between the choice of twisted/ordinary commutation relations and the huge rational functions in $q_{1},q_{2}$ on the RHS.

\begin{figure}[htb]
\centering
\includegraphics[scale=0.8]{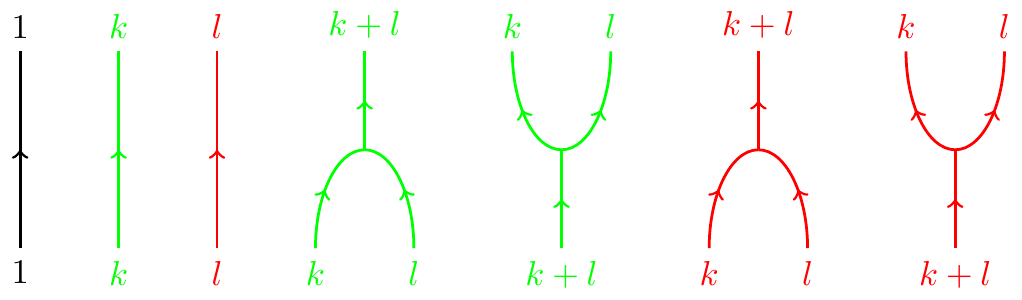}
\includegraphics[scale=0.8]{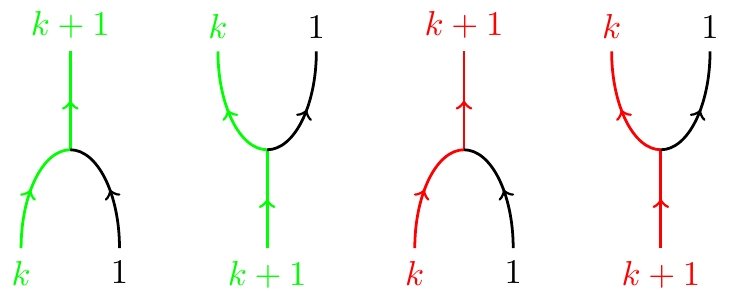}
\caption{Junctions that appear in the super Howe duality functor. Above: monochromatic edges and their trivalent junctions. Below: mixed-color trivalent junctions. Mirror images are also generators.}
\label{fig:superHoweGenerators}
\end{figure}

\section{Refined super Howe duality and $\beta$-deformed quantum supergroup}
\label{sec:super}

\begin{figure} [tbp]
\centering
\includegraphics{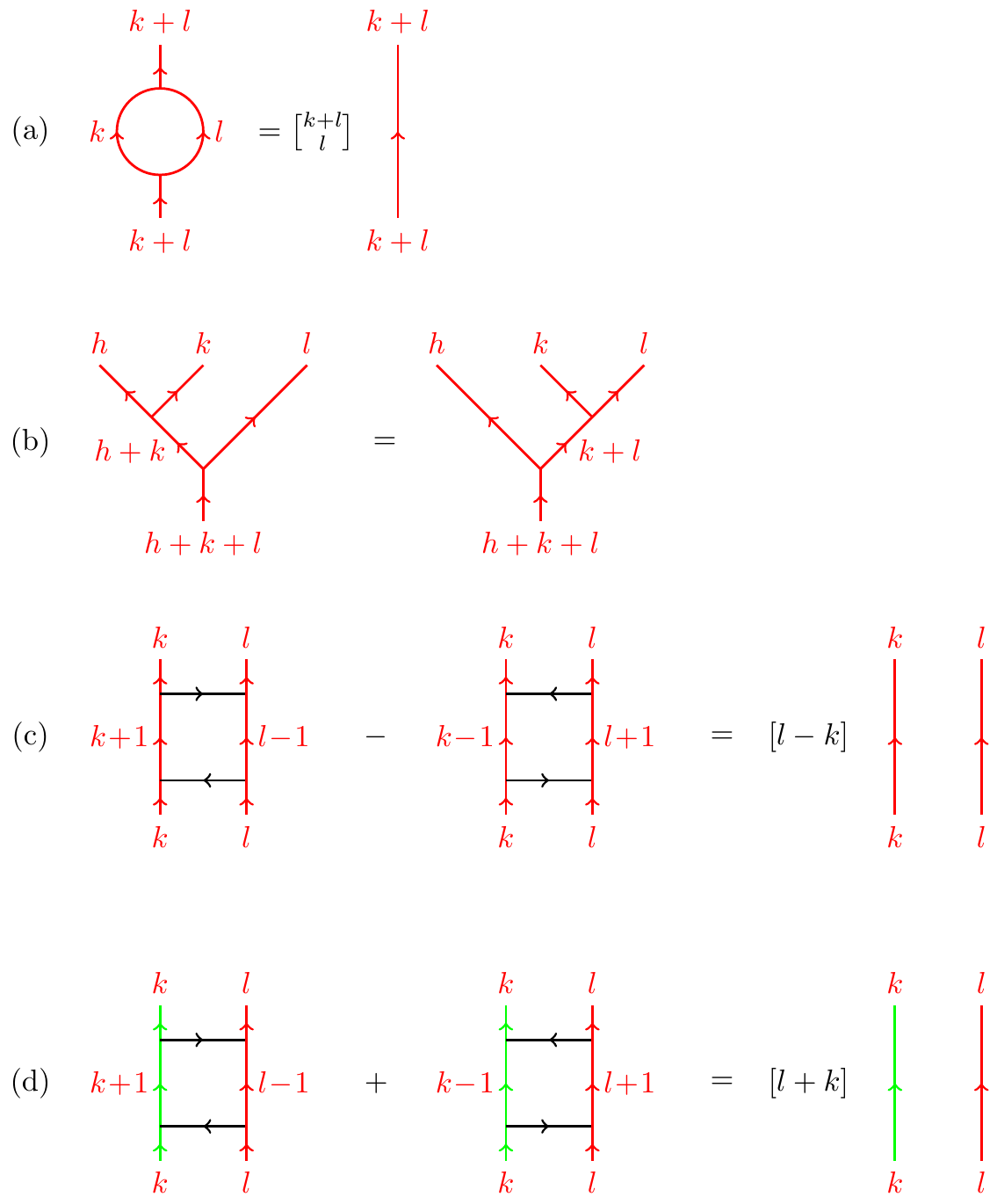}
\caption{Monochromatic relations (the same holds for green edges): (a) digon removal, (b) associativity, and (c) the monochromatic [E,F] relation. Mixed-color relation: (d) the mixed-color [E,F] relation.}
\label{fig:superHoweRelations}
\end{figure}

In this section, we include refined Wilson lines which are colored in symmetric representations and propose a refined super Howe duality. Following \cite{TVW}, we distinguish Wilson lines colored in symmetric/antisymmetric representations by their colors, as depicted in Figure \ref{fig:superHoweGenerators}. The red/green/black edges are colored in symmetric/antisymmetric/fundamental representations, respectively. The junctions in Figure \ref{fig:superHoweGenerators} generate the green-red web category of \cite{TVW}, and they satisfy the relations in Figure \ref{fig:superHoweRelations}. 

\subsection{Symmetric commutation relation}
\begin{figure} [htb]
\centering
\raisebox{-0.5\height}{\includegraphics{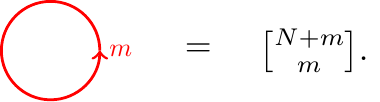}} \quad
\raisebox{-0.5\height}{\includegraphics{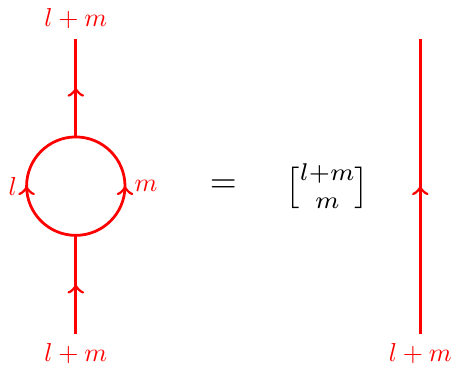}}
\caption{Left: unrefined expectation value of a Wilson loop in representation $Sym^{m}\square$. Right: unrefined digon removal relation of symmetric-colored Wilson lines.}
\label{fig:ord_circ_digon}
\end{figure}

To refine the network of Wilson lines colored in symmetric representations, we again consider the circle removal and the digon removal relations (Figure \ref{fig:ord_circ_digon}).

\begin{figure} [htb]
\centering
\raisebox{-0.5\height}{\includegraphics{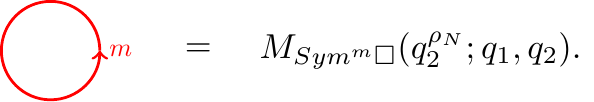}} \quad
\raisebox{-0.5\height}{\includegraphics{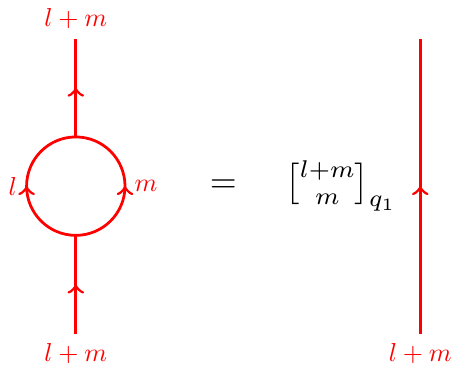}}
\caption{Left: refined expectation value of a Wilson loop in representation $Sym^{m}\square$. Right: refined digon removal relation of symmetric-colored Wilson lines.}
\label{fig:ref_circ_digon}
\end{figure}

The circle removal relation represents the expectation value of a Wilson loop colored in $Sym^{m}\square$, so it must be refined to Macdonald polynomials for $Sym^{m}\square$. The digon removal relation also allows an interpretation as a circle removal relation of a smaller gauge group, via an analogue of the level-rank duality: 
\begin{align*}
M_{Sym_{m}\square}(q_{2}^{\rho_{N}};q_{1},q_{2}) &= \dfrac{q_{2}^{\tfrac{N}{2}}-q_{2}^{-\tfrac{N}{2}}}{q_{2}^{\tfrac{1}{2}}-q_{2}^{-\tfrac{1}{2}}} \cdot \dfrac{q_{2}^{\tfrac{N}{2}}q_{1}^{\tfrac{1}{2}}-q_{2}^{-\tfrac{N}{2}}q_{1}^{-\tfrac{1}{2}}}{q_{2}^{\tfrac{1}{2}}q_{1}^{\tfrac{1}{2}}-q_{2}^{-\tfrac{1}{2}}q_{1}^{-\tfrac{1}{2}}} \cdots \dfrac{q_{2}^{\tfrac{N}{2}}q_{1}^{\tfrac{m-1}{2}}-q_{2}^{-\tfrac{N}{2}}q_{1}^{-\tfrac{m-1}{2}}}{q_{2}^{\tfrac{1}{2}}q_{1}^{\tfrac{m-1}{2}}-q_{2}^{-\tfrac{1}{2}}q_{1}^{-\tfrac{m-1}{2}}} \\[1.5ex]
&= \dfrac{q_{1}^{\tfrac{k}{2}}-q_{1}^{-\tfrac{k}{2}}}{q_{2}^{\tfrac{1}{2}}-q_{2}^{-\tfrac{1}{2}}} \cdot \dfrac{q_{1}^{\tfrac{k}{2}}q_{1}^{-\tfrac{1}{2}}-q_{1}^{-\tfrac{k}{2}}q_{1}^{\tfrac{1}{2}}}{q_{2}^{\tfrac{1}{2}}q_{1}^{\tfrac{1}{2}}-q_{2}^{-\tfrac{1}{2}}q_{1}^{-\tfrac{1}{2}}} \cdots \dfrac{q_{1}^{\tfrac{k}{2}}q_{1}^{-\tfrac{m-1}{2}}-q_{1}^{-\tfrac{k}{2}}q_{1}^{\tfrac{m-1}{2}}}{q_{2}^{\tfrac{1}{2}}q_{1}^{\tfrac{m-1}{2}}-q_{2}^{-\tfrac{1}{2}}q_{1}^{-\tfrac{m-1}{2}}} \\[1.5ex]
&= \quantumbinomial{k}{m}_{q_{1}} \cdot \dfrac{[m]_{q_{1}}!}{[m]_{*}!} = M_{\wedge^{m}\square}(q_{1}^{\rho_{k}};q_{2},q_{1}) \dfrac{[m]_{q_{1}}!}{[m]_{*}!},
\end{align*}
where the second equality follows from $q_{2}^{N/2} = e^{\pi i \beta N / (\beta N + k)} = - e^{-\pi i k/ (\beta N + k)} = - q_{1}^{-k/2}$. Notice that the correspondence between $M_{Sym_{m}\square}(q_{2}^{\rho_{N}};q_{1},q_{2})$ and $M_{\wedge^{m}\square}(q_{1}^{\rho_{k}};q_{2},q_{1})$ very much resembles the level-rank duality in ordinary Chern-Simons theory; in the refined case, replacing $\beta \leftrightarrow 1/\beta$ exchanges $q_{1} \leftrightarrow q_{2}$, $k \leftrightarrow N$ and $Sym^{m}\square \leftrightarrow \wedge^{m}\square$. Although such a correspondence is only an analogue, we can use it as an intuition to propose a digon removal relation for the symmetric Wilson lines (Figure \ref{fig:ref_circ_digon}).

Now let us study $[E,F]$ relation of symmetric-colored Wilson lines. We embed the Wilson lines of the symmetric $[E,F]$ relation into the Wilson lines of Figure \ref{fig:EF_close}, but this time the Wilson lines are colored in symmetric representations. The refined Littlewood-Richardson coefficients are trivial $\hat{N}_{\square,Sym^{l}\square}^{Sym^{l+1}\square} = 1 \, \forall l,$
so the homotopies do not involve global corrections. As a result,  we obtain the following linear relations:

\begin{gather}
\alpha [j+1]_{q_{1}}[m]_{q_{1}}M_{j+1}^{N} M_{m}^{N} + \beta [j]_{q_{1}}[m+1]_{q_{1}} M_{j}^{N} M_{m+1}^{N} + \gamma M_{1}^{N} M_{k}^{N} M_{l}^{N} = 0, \\
\alpha [j+1]_{q_{1}}[m]_{q_{1}} + \beta [j]_{q_{1}}[m+1]_{q_{1}} + \gamma = 0 \\[1.5ex]
\Rightarrow \alpha : \beta : \gamma = -\dfrac{[j+1]_{s}}{[j+1]_{q_{1}}} : \dfrac{[m+1]_{s}}{[m+1]_{q_{1}}} : [m-j]_{q_{1}}.
\label{eqn:symEF}
\end{gather}

Notice that Equation \ref{eqn:symEF} is related to the antisymmetric $[E,F]$ relation (Equation \ref{eqn:antiEF}) via $q_{1} \leftrightarrow q_{2}$. Therefore, the symmetric refined Serre relation can be obtained by exchanging $q_{1} \leftrightarrow q_{2}$ from its antisymmetric counterpart.
\begin{figure} [htb]
\centering
\includegraphics{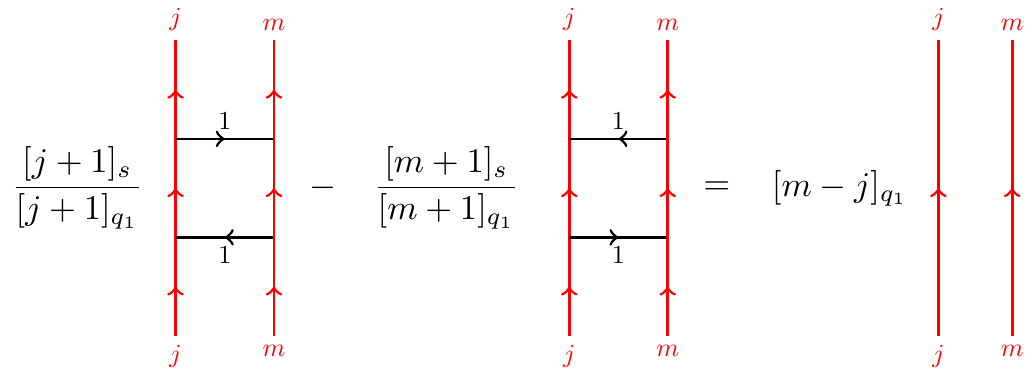} 
\caption{Above: symmetric, refined $[E,F]$ relation of junctions. Middle: a refinement of symmetric Howe duality functor. Below: refined $[E,F]$ relation rescaled to an ordinary commutation relation.}
\label{fig:symEF}
\end{figure}

\begin{figure} [htb]
\centering
\includegraphics{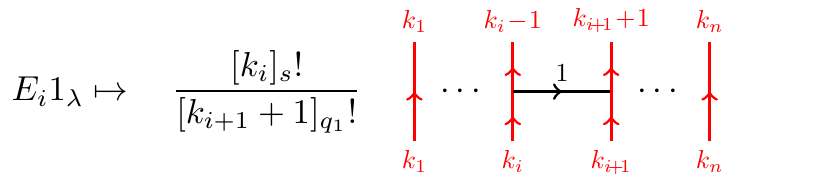}
\includegraphics{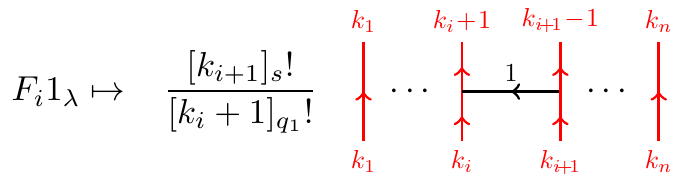} \\[1.5ex]
$[E_{i},F_{i}]1_{\lambda} = [k_{i+1}-k_{i}]_{q_{1}}\dfrac{[k_{i}]_{s}![k_{i+1}]_{s}!}{[k_{i}]_{q_{1}}![k_{i+1}]_{q_{1}}!}1_{\lambda}.$
\caption{A refinement of symmetric Howe duality functor, and the resultant symmetric $[E,F]$ commutation relation.}
\label{fig:symRescale}
\end{figure}

The symmetric $[E,F]$ relation of junctions in Figure \ref{fig:symEF} is again a twisted commutation relation. Therefore, we may attempt at finding a refinement of the symmetric Howe duality functor, which will turn the commutation relation of quantum group generators into a  commutation relation. The rescaling and the resultant non-twisted commutation relation is provided in Figure \ref{fig:symRescale}.

\subsection{Mixed-color commutation relation}
\begin{figure} [htb]
\centering
\includegraphics{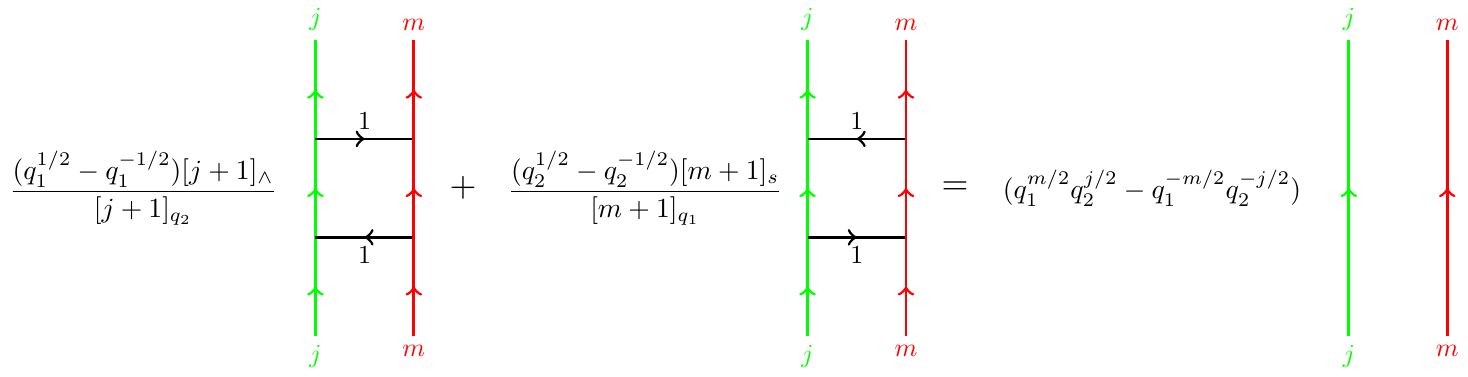}
\caption{Mixed-color $[E,F]$ relation of Wilson lines}
\label{fig:mixedEF}
\end{figure}

Of all the generating relations of the quantum supergroup, what remains is the mixed-color $[E,F]$ relation, Figure \ref{fig:superHoweRelations} (d). Following the derivation of \cite{TVW}, we can fix the coefficients of the relevant Wilson lines as in Figure \ref{fig:mixedEF}.

\begin{figure} [htb]
\centering
\includegraphics{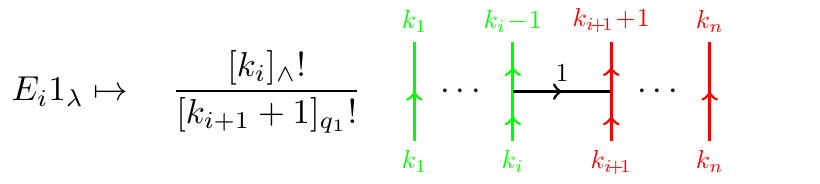}
\includegraphics{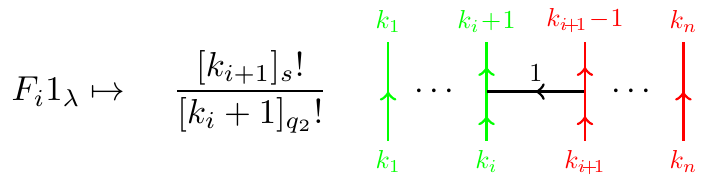}
\caption{A refinement of super Howe duality functor.}
\label{fig:mixedRescale}
\end{figure}

We may attempt at finding a refinement of the super Howe duality functor, by which the mixed-color $[E,F]$ relation becomes an untwisted anticommutation relation. It is indeed possible to do so, but such a refinement involves nontrivial powers of $(q_{1}^{1/2}-q_{1}^{-1/2})$ and $(q_{2}^{1/2}-q_{2}^{-1/2})$ on the RHS of the commuation relation, which totally obscures the connection with unrefined limit $q_{1} = q_{2}$. So instead we refine the super Howe duality functor as in Figure \ref{fig:mixedRescale}, and obtain the resultant quantum supergroup relation, which is a twisted commutator relation (but a more intuitive one.)

$$(q_{1}^{1/2}-q_{1}^{-1/2})EF1_{\lambda} + (q_{2}^{1/2}-q_{2}^{-1/2})FE1_{\lambda} = (q_{1}^{k_{i+1}/2}q_{2}^{k_{i}/2}-q_{1}^{-k_{i+1}/2}q_{2}^{-k_{i}/2})\dfrac{[k_{i+1}]_{s}![k_{i}]_{\wedge}!}{[k_{i+1}]_{q_{1}}[k_{i}]_{q_{2}}}1_{\lambda}.$$

\section{Discussions}
\begin{figure}[htb]
\centering
\includegraphics{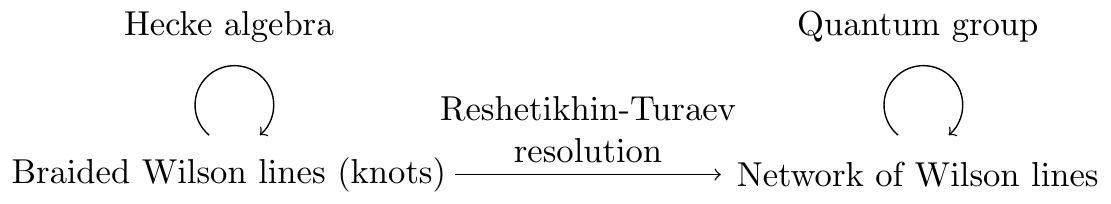}
\caption{Correspondence between the braided/network of Wilson lines}
\label{fig:diagram}
\end{figure}

As was pointed out multiple times throughout this paper, we need to understand the physical interpretation of junction degrees of freedom in the refined theory. Another question is whether we can compute the refined expectation values of knots/links by resolution of crossings. Recall that $\mathfrak{sl}_{N}$ polynomials can be computed by resolving the knots/links into a linear sum of MOY graphs. An analogous technique in the refined theory would allow us to compute the refined expectation values of non-torus knots/links as well. In the context of representation theory, Wilson lines with crossings come with a natural braid group action which factorizes through a Hecke algebra action. Thus, the polynomial knot invariants may be thought as polynomial representations of Hecke algebras. Now, let us replace each crossing by a linear sum of networks. Then, one finds the quantum group action on the resultant trivalent graphs as we have seen before. The question is whether the correspondence between Hecke algebra and quantum groups (summarized in Figure \ref{fig:diagram}) extend to their $\beta$-deformations.
Indeed, the expectation values of unknotted Wislon loops in refined Chern-Simons theory are given by MacDonald polynomials, which are polynomial representations of double affine Hecke algebra (DAHA). If the correspondence extends to the refined case, our $\beta$-deformed quantum groups can serve as the starting point for the higher representation theory of DAHA, much of which is unknown at present.

\acknowledgments{The author is deeply indebted to Aaron Lauda and Sergei Gukov for their suggestions and invaluable discussions.

The work is funded in part by the DOE Grant DE-SC0011632 and the Walter Burke Institute for Theoretical Physics, and also by the Samsung Scholarship.}

\appendix
\section{A no-go theorem: the refined commutation relation of junctions cannot be put into a categorification-friendly commutation relation of quantum group generators}
\label{appen:digon}
In this section, we exhibit that the refined $[E,F]$ relation cannot be written in a categorification-friendly commutation relation. Let us first write down $[E,F]$ and Serre relation of antisymmetric colored Wilson lines, with the digon removal relation left as abstract. The purpose of such an abstraction is to exhibit that the ominous rational functions of $q_{1},q_{2}$ on the RHS of Equation \ref{eqn:omg} are indeed inevitable.  
\begin{figure} [htb]
\centering
\includegraphics{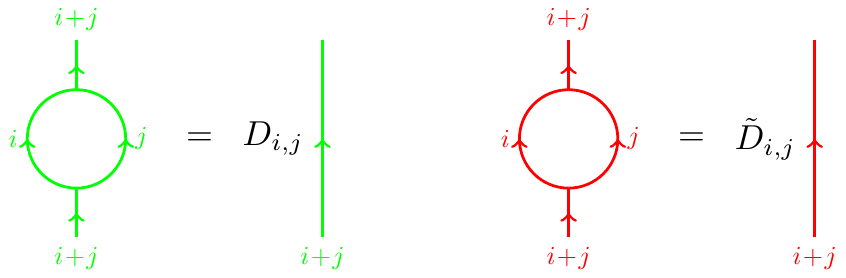}
\caption{An abstract digon removal relation. Left: digon removal relation for antisymmetric-colored Wilson lines. Right: digon removal relation for the symmetric-colored Wilson lines.}
\label{fig:abstract_digon}
\end{figure}

Using the abstract digon removal relation depicted in Figure \ref{fig:abstract_digon}, the Equation \ref{eqn:antiEF} and Equation \ref{eqn:symEF} becomes:
\begin{gather}
\alpha_{j,m}N_{1,j}^{1+j}D_{1,j}D_{1,m-1}M_{j+1}M_{m} + \beta_{j,m}N_{1,m}^{1+m}D_{1,j-1}D_{1,m}M_{j}M_{m+1} + \gamma_{j,m}M_{1}M_{j}M_{m} = 0, \\
\alpha_{j,m}D_{1,j}D{1,m-1} + \beta_{j,m}D_{1,j-1}D_{1,m} + \gamma = 0 \\[1.5ex]
\Rightarrow \alpha_{j,m}D_{1,j}D_{1,m-1} : \beta_{j,m}D_{1,j-1}D_{1,m} : \gamma = -[j+1]_{\wedge}[m]_{q_{2}} : [j]_{q_{2}}[m+1]_{\wedge} : [m-j]_{q_{2}}
\end{gather} for the antisymmetric-colored Wilson lines. Then, the refined $[E,F]$ relation becomes (Figure \ref{fig:EF_abst}):
\begin{figure} [htb]
\centering
\includegraphics{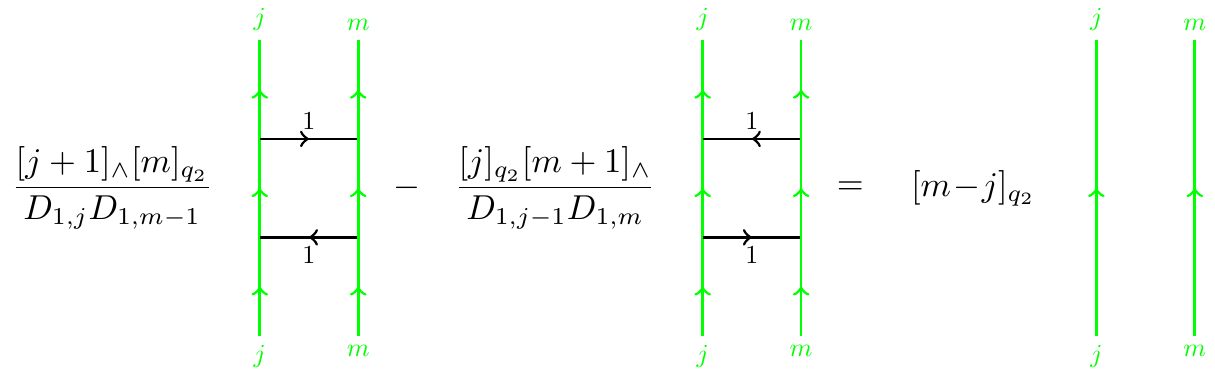}
\caption{Refined antisymmetric $[E,F]$ relation with digon removal abstraction}
\label{fig:EF_abst}
\end{figure}

Is there a smart choice of $D_{i,j}$ which turns Figure \ref{fig:EF_abst} into an ordinary (non-twisted) $[E,F]$ relation? Let's suppose we can, and then $D_{i,j}$'s must satisfy the following equation:

\begin{gather}
\frac{D_{1,j}}{D_{1,j-1}}\frac{D_{1,m-1}}{D_{1,m}} = \frac{[j+1]_{\wedge}}{[j]_{q_{2}}}\frac{[m]_{q_{2}}}{[m+1]_{\wedge}} \\
\Rightarrow \quad \frac{D_{1,j}}{D_{1,j-1}} = \frac{[j+1]_{\wedge}}{[j]_{q_{2}}}.
\label{eqn:rescale_undet}
\end{gather}
The last line follows from the fact that the $j$- and $m$-dependence factorize. One choice of $D_{1,j}$ which allows us to obtain the ordinary commutation relation is:
$$D_{1,j} = [j+1]_{\wedge}!/[j]_{q_{2}}!$$

Recall that $D_{1,j}$, the coefficient of a local digon removal relation, counts the graded dimension of a composite morphism $\wedge^{1+j}\square \rightarrow \square \otimes \wedge^{j}\square \rightarrow \wedge^{1+j}$. The above expression, although it reduces to $[j+1]_{q}$ in the unrefined limit, does not seem to be so categorification-friendly. What if we add the rescaling of quantum group generators? Can we get an ordinary commutation relation which is also categorification-friendly? Combining Figure \ref{fig:rescale} and Figure \ref{fig:EF_abst}, we find the following refined $[E,F]$ relation of quantum generators:
\begin{equation}
\frac{[j+1]_{\wedge}[m]_{q_{2}}}{D_{1,j}D_{1,m-1}}\frac{1}{C_{j+1,m-1}C'_{j,m}}EF1_{\lambda} - \frac{[j]_{q_{2}}[m+1]_{\wedge}}{D_{1,j-1}D_{1,m}}FE1_{\lambda}\frac{1}{C_{j,m}C'_{j-1,m+1}} = [m-j]_{q_{2}}1_{\lambda}.
\label{eqn:antiEF_fullabst}
\end{equation}

Suppose we can put Equation \ref{eqn:antiEF_fullabst} into an ordinary commutation relation. Then, $C_{j,m},C'_{j,m}$ and $D_{i,j}$ must satisfy the following equation:

\begin{equation}
\frac{C_{j+1,m-1}}{C_{j,m}}\frac{C'_{j,m}}{C'_{j-1,m+1}} = \frac{[j+1]_{\wedge}D_{1,j-1}}{[j]_{q_{2}}D_{1,j}}\frac{[m]_{q_{2}}D_{1,m}}{[m+1]_{\wedge}D_{1,m-1}},
\label{eqn:antiEF_fact1}
\end{equation}
where again the $j$- and $m$-dependence factorizes. Therefore, we may factorize $C_{j,m}$ and $C'_{j,m}$ into $C_{j,m} = A_{j}B_{m}, \, C'_{j,m} = A'_{j}B'_{m}$. Then, the Equation \ref{eqn:antiEF_fact1} becomes:

\begin{align}
\frac{A_{j+1}A'_{j}}{A_{j}A'_{j-1}} &= \frac{[j+1]_{\wedge}D_{1,j-1}}{[j]_{q_{2}}D_{1,j}}, \quad  \frac{B_{m-1}B'_{m}}{B_{m}B'_{m+1}} = \frac{[m]_{q_{2}}D_{1,m}}{[m+1]_{\wedge}D_{1,m-1}}, \\[1.5ex]
\Rightarrow \quad &A_{j+1}A'_{j} = \frac{[j+1]_{\wedge}D_{1,j-1}}{[j]_{q_{2}}D_{1,j}} \frac{[j]_{\wedge}D_{1,j-2}}{[j-1]_{q_{2}}D_{1,j-1}} \cdots \frac{[2]_{\wedge}D_{1,0}}{[1]_{q_{2}}D_{1,1}} A_{1}A'_{0} = \frac{[j+1]_{\wedge}!}{[j]_{q_{2}}! D_{1,j}} A_{1}A'_{0}, \\[1.5ex]
&B_{m}B'_{m+1} = \frac{[m+1]_{\wedge}D_{1,m-1}}{[m]_{q_{2}}D_{1,m}} \frac{[m]_{\wedge}D_{1,m-2}}{[m-1]_{q_{2}}D_{1,m-1}} \cdots \frac{[2]_{\wedge}D_{1,0}}{[1]_{q_{2}}D_{1,1}} B_{0}B'_{1} = \frac{[m+1]_{\wedge}!}{[m]_{q_{2}}! D_{1,m}} B_{0}B'_{1}.
\label{eqn:antiEF_fact2}
\end{align}

Now notice that $A_{1}B_{0} = A'_{0}B'_{1} = 1$, for $E1_{(1,0)}$ and $F1_{(0,1)}$ are homotopic to identity morphisms. Therefore, we may simply put $A_{1} = A'_{0} = A_{1} = B_{0} = 1$ (since they possess no $j$- or $m$-dependence) so that the last two equalities of Equation \ref{eqn:antiEF_fact2} becomes:

\begin{gather}
A_{j+1}A'_{j} = \frac{[j+1]_{\wedge}!}{[j]_{q_{2}}!D_{1,j}}, \quad B_{m}B'_{m+1} = \frac{[m+1]_{\wedge}!}{[m]_{q_{2}}! D_{1,m}} \\[1.5ex]
\Rightarrow C_{j,m}C'_{j-1,m+1} = \frac{[j]_{q_{2}}[m+1]_{\wedge}}{D_{1,j-1}D_{1,m}}\frac{[j]_{\wedge}![m]_{\wedge}!}{[j]_{q_{2}}![m]_{q_{2}}!}.
\label{eqn:ABA'B'}
\end{gather} 

Consider any choice of $A_{j},B_{m},A'_{j},B'_{m}$ which satisfy Equation \ref{eqn:ABA'B'}. Plugging them in the Equation \ref{eqn:antiEF_fullabst}, we can rescale the refined $[E,F]$ relation to an ordinary commutation relation as follows: 
\begin{equation}
[E,F]1_{\lambda} = [m-j]_{q_{2}}\frac{[j]_{\wedge}![m]_{\wedge}!}{[j]_{q_{2}}![m]_{q_{2}}!}1_{\lambda}.
\label{eqn:monster}
\end{equation}
This shows that once we rescale the refined $[E,F]$ relation to an ordinary commutation relation, the monstrous multiplication coefficient on the RHS of Equation \ref{eqn:monster}. For symmetric-colored Wilson lines the situation is completely analogous, and we only need to exchange $q_{1} \leftrightarrow q_{2}$.

\newpage

\bibliographystyle{JHEP_TD}
\bibliography{RefinedWeb}

\end{document}